\documentclass[11pt]{article}
\usepackage{fullpage}
\usepackage{amsmath,amssymb,graphicx,amsthm,dsfont}
\usepackage{color,soul}
\usepackage{xspace}
\usepackage{enumerate}
\usepackage{expdlist}
\usepackage{mathtools}
\usepackage{bbm}
\usepackage{thm-restate}
\usepackage{url}

\usepackage{multicol}
\usepackage{booktabs}
\usepackage[square]{natbib}
\usepackage{float}
\usepackage{graphicx}
\usepackage{tikz-cd}
\usepackage{tikz}
\usetikzlibrary{positioning,chains,fit,shapes,calc,snakes}
\usepackage[normalem]{ulem}
\usepackage[ruled]{algorithm2e} 
\usepackage{cleveref}
\usepackage{xcolor}

\newcommand{\rep}{\ensuremath{\text{rep}}}
\newcommand{\h}{\ensuremath{\mathbf{h}}}
\newcommand{\x}{\ensuremath{\mathbf{x}}}
\newcommand{\y}{\ensuremath{\mathbf{y}}}
\newcommand{\rr}{\ensuremath{\mathbf{r}}}
\newcommand{\rank}{\ensuremath{\text{rank}}}
\newcommand{\profile}{\boldsymbol{\succ}}

\newcommand{\LP}{\ensuremath{\text{LP}}}

\newtheorem{theorem}{Theorem}

\newtheorem{definition}[theorem]{Definition}

\newtheorem*{example}{Example}

\newtheorem{question}{Open Problem}

\newcommand{\e}{\ensuremath{\mathbf{e}}}

\title{Quantile agent utility and implications to randomized social choice}
\author{Ioannis Caragiannis\thanks{Department of Computer Science, Aarhus University, Denmark. Email: iannis@cs.au.dk.} \and Fabian Frank \thanks{Technical University of Munich} \and Sanjukta Roy\thanks{Indian Statistical Institute, Kolkata, India. Email: sanjukta@isical.ac.in} }
\date{}

\begin{document}

\maketitle

\begin{abstract}
We initiate a novel direction in randomized social choice by proposing a new definition of agent utility for randomized outcomes. Each agent has a preference over all outcomes and a {\em quantile} parameter. Given a {\em lottery} over the outcomes, an agent gets utility from a particular {\em representative}, defined as the least preferred outcome that can be realized so that the probability that any worse-ranked outcome can be realized is at most the agent's quantile value. 

In contrast to other utility models that have been considered in randomized social choice (e.g., stochastic dominance, expected utility), our {\em quantile agent utility} compares two lotteries for an agent by just comparing the representatives, as is done for deterministic outcomes. This yields a purely ordinal yet informative comparison of randomized outcomes.

We revisit fundamental questions in randomized social choice using the new utility definition. We study the compatibility of efficiency and strategyproofness for randomized voting rules, and of efficiency, fairness, and strategyproofness for randomized one-sided matching mechanisms. In contrast to classical impossibility results, we show that under quantile agent utilities, these properties can often be satisfied simultaneously.
\end{abstract}

\section{Introduction}
{\em Randomized mechanisms} play a central role in social choice and matching theory, offering a powerful tool for navigating fundamental tensions between fairness, efficiency, and incentives. Their importance has become increasingly visible in real-world applications recently. For example, during the COVID-19 pandemic, randomized procedures were advocated for vaccine allocation to ensure equitable access across diverse populations~\citep{NEJM20,White22}. More broadly, randomization is routinely used today to allocate scarce public resources—such as school seats, hospital beds, and public housing—in a way that balances fairness and efficiency. In voting and collective decision-making, randomized mechanisms are used to break ties or to select representatives fairly in multi-agent decision-making environments, such as in citizen assemblies~\citep{Flanigan2021FairAssemblies}, where no deterministic method can satisfy all normative requirements.

Despite this promise, randomized mechanisms continue to face fundamental limitations. Classical negative results show that even when randomization is allowed, achieving key axioms such as efficiency and strategyproofness simultaneously can be {\em impossible}. In voting, for example, Gibbard’s theorem implies that randomized strategyproof and efficient rules must be (randomized versions of) dictatorships~\citep{G77chance}.

However, such negative results strongly depend on the assumptions regarding how voters evaluate randomized outcomes. In much of the literature, agents are assumed to compare lotteries either by positing underlying cardinal utilities and using {\em expected utility}, or by adopting fully ordinal dominance criteria such as {\em stochastic dominance}. Although expected utility induces a complete comparison over lotteries, it relies on cardinal information that is often viewed as arbitrary or unrealistic in social choice settings. Stochastic dominance, on the other hand, avoids cardinal assumptions but is a strong requirement that can render many lotteries {\em incomparable}, thereby reintroducing impossibilities even in randomized environments~\citep{a18,B17}.

More specifically, consider a classical voting scenario, in which one would like to aggregate preferences over alternatives submitted by the voters (or agents) into a single winning alternative. An agent's preference (usually, a strict ranking) over the alternatives determines how the agent evaluates different deterministic outcomes that may be returned by a voting rule and compares them to each other. But how do agents evaluate the outcome of randomized social choice procedures such as randomized voting rules? A randomized voting rule takes as input a profile of agent preferences (rankings over the alternatives) and returns a random alternative according to a probability distribution or, in other words, a {\em lottery} over the alternatives. 

The important question now is: how can an agent decide which among two lotteries is better using her preference ranking over the alternatives? The answer is no longer trivial. For example, how can an agent with preference $a\succ b \succ c$ over three alternatives $a$, $b$, and $c$ compare the lottery that returns equiprobably the three alternatives with the lottery that returns alternative $b$ with probability $1$? There have been several answers in the literature. One approach is to assume that the agent has a valuation function that assigns a cardinal utility for each alternative that respects the preference relation of the agent for the alternatives. In our example, the utility for alternative $a$ has to be higher than that for alternative $b$, which in turn has to be higher than that for alternative $c$. The two lotteries can now be compared according to the expected utility of the agent for the alternative they return. This approach requires the existence of underlying cardinal utilities, which the agents should be able to compute; this is not always realistic.

Another prominent approach in randomized social choice is to compare lotteries using the notion of stochastic dominance~\citep{B17}. According to it, a lottery is better than another for an agent if it yields higher expected utility for her for every possible underlying utilities for the alternatives that respect the agent's preference ranking.\footnote{We remark that this is a non-standard definition of stochastic dominance. The original definition is more technical and is given in Section~\ref{sec:concept}.} This is a very stringent definition which may result in two lotteries being incomparable. For example, the comparison between the two lotteries mentioned above gives a different outcome for the underlying utilities $(1, 0.2, 0)$ and $(1, 0.8, 0)$ for alternatives $a$, $b$, and $c$, respectively. The expected utilities for the two lotteries are $0.4$ and $0.2$ for the former utility values and $0.6$ and $0.8$ for the latter (see Table~\ref{tab:intro-example}).

\begin{table}[h]
\centering
\begin{tabular}{c|cc c}
utilities respecting & \multicolumn{2}{c}{lottery} &  \\
preference $a\succ b\succ c$ & $x=(1/3,1/3,1/3)$ & $y=(0,1,0)$ & comparison\\\hline
$(1, 0.2, 0)$ & \boxed{$0.4$} & $0.2$ & $x$ is better\\
$(1, 0.8, 0)$ & $0.6$ & \boxed{$0.8$} & $y$ is better
\end{tabular}

    \caption{Comparing lotteries $x$ and $y$ under different cardinal utilities consistent with the ordinal preference $a \succ b \succ c$. The table shows the expected values for the two lotteries $x$ and $y$ for the two underlying utilities for alternatives $a$, $b$, and $c$, showing they are incomparable under stochastic dominance.}
    \label{tab:intro-example}
\end{table}

Motivated by these limitations, we present the novel {\em quantile agent utility} model for the comparison of lotteries. Our framework shifts the focus away from cardinal utility assumptions and from overly demanding dominance criteria, and instead evaluates lotteries based on an agent’s least preferred outcome that can occur with sufficiently high probability. Our aim is to reconcile the often conflicting objectives of efficiency, fairness, and strategyproofness in social choice mechanisms.

More concretely, each agent has a {\em quantile parameter} (a scalar between $0$ and $1$), and associates each lottery over alternatives with the highest-ranked alternative that has the following properties: it has positive probability and it is such that the less preferred alternatives have total probability that does not exceed the quantile parameter. We use the term {\em representative} to refer to this alternative. Then, the comparison of two lotteries by an agent is simply the comparison of their representative alternatives according to the agent's preference. In decision-theoretic terms, the different quantile parameters can model agents with different risk levels in their interpretation of a lottery~\citep{manski1988ordinal}.

Our quantile agent utility model has several advantages. First, the comparison between lotteries can take place in the same way as in the case of deterministic outcomes. Second, as we will see, it turns out that stochastic dominance is as stringent in the comparison between two lotteries as our utility definition would be by considering {\em all} possible quantile parameters. And, of course, there are no underlying cardinal utilities for the alternatives that the agent needs to be able to evaluate the lotteries.

\subsection{Our contribution}
The definition of the quantile agent utility model is our main conceptual contribution. We define it in Section~\ref{sec:concept}, where we also demonstrate that it is considerably less stringent than stochastic dominance. Next, we explore its implications for two social choice settings: voting and one-sided matchings. Our central message is that many impossibility results in these settings are artifacts of overly strong lottery comparison models, and that by adopting quantile agent utilities, efficiency, fairness, and strategyproofness can often be reconciled.

Specifically, with quantile agent utilities, we get different results that depend on the range of quantile parameter values of the agents. For example, if all agents have their quantile parameter equal to $0$, we usually recover the impossibilities for deterministic mechanisms. If, instead, all quantile parameters are close to $1$, trivial mechanisms like uniform lotteries are usually ideal. Intermediate quantile parameter values, possibly different for each agent, allow for a spectrum of interesting results. 

\paragraph{\bf Voting.} In Section~\ref{sec:voting}, we consider the classical ranking-based voting scenario, in which a {\em voting rule} aims to select a winning alternative (candidate) taking as input rankings of the available alternatives that are submitted as votes by a set of agents. The voting rules are randomized, and the agents have quantile utilities. We adapt the definition of (Pareto) {\em efficiency} in this context and first observe that no lottery can be efficient for all vectors of quantile parameters for the agents unless all agents agree on their top-ranked alternative. This justifies our choice to focus on scenarios in which each agent has fixed their quantile parameter, which in turn can be used in the definition of the lottery returned by the voting rule. We then adapt the notion of {\em strategyproofness} and explore its relation to monotonicity in the new context for two alternatives. We present the randomized voting rule $Q$-plurality for two alternatives, which we prove to be efficient and strategyproof. We conclude the section with the more intricate case of three alternatives. For agents with relatively high quantile parameters, we construct explicit non-dictatorial voting rules that are efficient and strategyproof, thereby circumventing the classical impossibility result of~\citet{G77chance} for randomized voting under stochastic dominance; for small quantile parameters, we show that deterministic dictatorships are the only efficient and strategyproof rules. Extending our characterization for all possible quantile parameter vectors is left as an (apparently challenging) important open problem.

\paragraph{\bf One-sided matching.} In Section~\ref{sec:matching}, we consider scenarios with a set of agents that are to be matched (in a one-to-one manner) to a set of items of equal size. Each agent has a ranking of the items representing her preference over them and a quantile parameter. We consider randomized {\em matching mechanisms} that return lotteries over perfect matchings. These lotteries define a representative item for each agent; the position of this item in the agent's ranking indicates how desirable the lottery is for the agent. We begin by considering a simple variant of social welfare in this context and show that optimal lotteries over matchings can be easily computed. We also discuss a more elaborate definition of social welfare, which turns out to be NP-hard to optimize. Again, we adapt the notions of efficiency and strategyproofness and demonstrate how to adapt {\em serial dictatorship} in our context using linear programming. In contrast to a well-known characterization for efficient deterministic matchings due to~\citet{AS98}, we show that there are efficient matching lotteries that cannot be produced by serial dictatorship. Next, motivated by the fair division literature, we consider fairness properties. We define the analog of {\em proportionality} and show how to adapt serial dictatorships to get a strategyproof matching mechanism that computes efficient and proportional lotteries. 
Furthermore, we give a natural definition of {\em envy-freeness}. Our definitions guarantee that envy-freeness implies proportionality (similar to the classical fair division literature; e.g., see~\citealp{BT96}). We observe that envy-freeness may not be consistent with efficiency; whether envy-free and efficient lotteries can be computed in polynomial time is left as an open problem. Interestingly, a modified version of serial dictatorship is proved to be strategyproof, efficient, and envy-free as long as no two agents have the same quantile parameter. 

\subsection{Related work}
Quantile-based utility models originate in decision theory. \citet{manski1988ordinal} introduced quantile utilities to model individual decision-making under uncertainty without relying on cardinal utilities, and \citet{rostek2010quantile} studied such models in continuous settings with risk preferences. These papers focus on individual choice and equilibrium behavior rather than on collective decision-making. To the best of our knowledge, quantile utilities have not previously been applied to social choice or matching theory, nor used as a foundation for studying axiomatic properties of randomized mechanisms, such as efficiency, fairness, and strategyproofness.

Lotteries over alternatives were first formally studied by~\citet{zeckhauser1969majority}, \citet{fishburn1972lotteries}, and \citet{intriligator1973probabilistic}. In voting, randomization has long been explored as a means of circumventing classical impossibility results for deterministic mechanisms, such as those established by Arrow’s Impossibility Theorem~\citep{arrow1951} and the Gibbard–Satterthwaite theorem \citep{Gib73,Sat75}, which together imply that deterministic strategyproof voting rules must be dictatorial. This motivated the design of randomized voting rules with improved incentive and fairness properties. For example, \citet{bogomolnaia2001} proposed the Random Priority mechanism, and \citet{procaccia2010} designed randomized voting rules that approximate score-based deterministic rules while achieving strategyproofness. More recently, \citet{a18} showed that the compatibility of efficiency and strategyproofness in randomized social choice depends crucially on how ordinal preferences are extended to lotteries. They studied extensions based on stochastic dominance (SD), pairwise comparisons (PC), bilinear dominance (BD), and the sure-thing principle (ST), and established several impossibility and incompatibility results. See also the survey by~\citet{B17}.

In resource allocation, where items need to be allocated to agents, envy-freeness~\citep{foley1966resource,varian1973equity} and proportionality~\citep{steinhaus1948problem} have been widely studied as measures of fairness. Specifically for one-sided matching, a prominent example is the house allocation problem, in which agents have strict preferences over items. \citet{hylland1979efficient} studied randomized matching mechanisms that are envy-free and ex ante efficient. \citet{abdulkadirouglu2003ordinal} further developed the study of lotteries in the house allocation setting. More broadly, randomization has been extensively explored in resource allocation and fair division problems; see, e.g., \citet{aziz2019probabilistic} for a survey of benefits and challenges. Within this literature, \citet{caragiannis2021interim} studied interim envy-freeness for lotteries over matchings.

\section{The quantile agent utility model }\label{sec:prelim}\label{sec:concept}
Formally, we define the quantile utility model as follows. Let $O$ denote a finite set of {\em options}. Consider an agent equipped with a strict preference order $\succ$ among the options of $O$ and a quantile parameter $h\in [0,1)$. Given a {\em lottery} (or probability distribution) $x$ over the options in $O$, we call option $o\in O$ the $h$-{\em quantile representative}\footnote{We can define the $1$-quantile representative of an agent to be the highest-ranked option that has positive probability to be returned by the lottery. However, to keep the exposition simple and the proofs easy to follow, we have decided to restrict quantile parameters in $[0,1)$.} of the agent if 
\begin{align*}
    \sum_{o\succ o'}{x(o')} &\leq h < \sum_{o\succeq o'}{x(o')}.
\end{align*}
The $h$-quantile representative of the agent in lottery $x$ is denoted by $\rep(x,\succ,h)$. We often use ``representative'' instead of ``$h$-quantile representative'' if the quantile parameter is clear from context. Table~\ref{tab:example-rep} shows examples of representatives for different lotteries and quantile parameters under the preference order $a\succ b \succ c$ over the set of options $O=\{a,b,c\}$.

\begin{table}[h]
\begin{tabular}{c|ccccccc}
 & \multicolumn{7}{c}{quantile parameter} \\
lottery over $(a,b,c)$ & $0$ & $1/4$ & $1/3$ & $1/2$ & $2/3$ & $3/4$ & $0.99$\\\hline
$(0,1,0)$ & $b$ & $b$ & $b$ & $b$ & $b$ & $b$ & $b$ \\
$(1/4,1/2,1/4)$ & $c$ & $b$ & $b$ & $b$ & $b$ & $a$ & $a$ \\
$(1/2,1/4,1/4)$ & $c$ & $b$ & $b$ & $a$ & $a$ & $a$ & $a$ \\
$(1/3,1/3,1/3)$ & $c$ & $c$ & $b$ & $b$ & $a$ & $a$ & $a$ \\
$(1/3,0,2/3)$ & $c$ & $c$ & $c$ & $c$ & $a$ & $a$ & $a$ 
\end{tabular}
\caption{The $h$-quantile representative $\rep(x,\succ,h)$ of an agent with the preference $a\succ b\succ c$ over the set of options $O=\{a, b, c\}$ for different quantile parameters $h$ and lotteries $x=(x(a),x(b),x(c))$.}
\label{tab:example-rep}
\end{table}

Now, the agent can compare two lotteries by just comparing her representatives under them. Thus, she may prefer one lottery to the other or be indifferent between the two. For example, by inspecting Table~\ref{tab:example-rep}, we can see that the agent prefers lottery $(1/4,1/2,1/4)$ to lottery $(1/3,0,2/3)$ when her quantile parameter is $1/2$, prefers lottery $(1/3,0,2/3)$ to lottery $(1/4,1/2,1/4)$ when the quantile parameter is $2/3$, and is indifferent between the two when the quantile parameter is $3/4$ or higher. Formally, given two lotteries $x$ and $y$ over the options in $O$, we say that an agent with a preference $\succ$ over the options in $O$, prefers $x$ to $y$ if and only if $\rep(x,\succ,h)\succ \rep(y,\succ,h)$. We say that the agent weakly prefers $x$ to $y$ if and only if $\rep(x,\succ,h)\succeq \rep(y,\succ,h)$ (i.e., either $\rep(x,\succ,h)\succ \rep(y,\succ,h)$ or $\rep(x,\succ,h)= \rep(y,\succ,h)$). 

We now explore the connection of our lottery extensions to the notion of {\em stochastic dominance}; we use the definition of~\citet{B17}. 
\begin{definition}[stochastic dominance]
We say that lottery $x$ over the set of options $O$ {\em stochastically dominates} lottery $y$ with respect to a preference $\succ$ over the options (and write $x\succsim^{\text{sd}} y$) if and only if 
\begin{align*}
    \sum_{o':o'\succeq o}{x(o')} &\geq \sum_{o':o'\succeq o}{y(o')}
\end{align*}
for every option $o\in O$. 
\end{definition}
We say that lottery $x$ strictly stochastically dominates lottery $y$ when we have $x\succsim^{\text{sd}}y$ but not $y\succsim^{\text{sd}}x$. Stochastic dominance and its strict version provide a way to compare lotteries over options when a preference over the options is available. For instance, in our stylized example, lottery $(3/4,0,1/4)$ stochastically dominates lottery $(1/2,0,1/2)$ and lottery $(1/2,1/4,1/4)$ stochastically dominates lottery $(1/4,1/2,1/4)$. Both relations are strict.

However, how do the lotteries $(1/2,0,1/2)$ and $(1/3,1/3,1/3)$ compare to each other? Notice that the former has more probability mass on the top option $a$ ($1/2$ vs.~$1/3$) while the latter has more probability mass on the two top options $a$ and $b$ ($2/3$ vs.~$1/2$). Hence, according to stochastic dominance, these two lotteries are incomparable. We remark that this inability for an agent to compare is very different than being indifferent between the two lotteries. Even though we do not attempt to define any axiom (like efficiency or strategyproofness) here, it must be clear that such incomparabilities play a crucial role in randomized social choice that uses stochastic dominance in the definition of these axioms.

In the example discussed in Table~\ref{tab:intro-example}, we demonstrated the incomparability of stochastic dominance via an equivalent definition that relates it to expected utility. It is well-known (e.g., see~\citealt{Sen1970CCSW}) that a lottery stochastically dominates another if it has at least as high expected utility for {\em all} underlying cardinal utilities for the options that respect the preference order. In a similar vein, we show that stochastic dominance is much more stringent than quantile agent utility as well. The proof of the next statement follows by carefully using the definitions of quantile agent utility and stochastic dominance.

\begin{restatable}{theorem}{thmstringent}\label{thm:stringent}
    Let $O$ be a set of options, $x$ and $y$ be lotteries over the options in $O$, and $\succ$ be a preference order over $O$. Then, $x\succsim^{\text{sd}} y$ if and only if $\rep(x,\succ,h)\succeq \rep(y,\succ,h)$ for every $h\in [0,1)$.
\end{restatable}

\begin{proof}
    First, assume that $x\succsim^{\text{sd}} y$ and, for the sake of contradiction, $a=\rep(y,\mathord{\succ},h^*)\succ \rep(x,\mathord{\succ},h^*)=b$ for some $h^*\in [0,1)$. By the definition of the $h^*$-quantile representative $\rep(x,\mathord{\succ},h^*)$, we get 
    \begin{align}\label{eq:def-rep-x}
        \sum_{o':b\succeq o'}{x(o')} > h^*.
    \end{align}
    Furthermore, 
    \begin{align}\label{eq:def-rep-y-a-b}
        h^* &\geq \sum_{o':a\succ o'}{y(o')}\geq \sum_{o':b\succeq o'}{y(o')}.
    \end{align}
    In this last equation, the first inequality follows from the definition of the $h^*$-quantile representative and the second one follows from the assumption that $a \succ b$.
    By equations (\ref{eq:def-rep-x}) and (\ref{eq:def-rep-y-a-b}), we get $\sum_{o':b\succeq o'}{x(o')} > \sum_{o':b\succeq o'}{y(o')}$. Denoting by $c$ the least preferred option in $O$ that satisfies $c\succ b$, we get $\sum_{o':o'\succeq c}{x(o')}<\sum_{o':o'\succeq c}{y(o')}$, contradicting the assumption $x\succsim^{\text{sd}} y$. This completes the proof of the ``only if'' part of the statement in Theorem~\ref{thm:stringent}.

    Now, assume that $\rep(x,\succ,h)\succeq \rep(y,\succ,h)$ for every $h\in [0,1)$. For the sake of contradiction, assume that $x\not\succsim^{\text{sd}} y$. Then, there exist options $o,a\in O$ that are consecutive under order $\succ$ satisfying $a\succ o$, so that 
    \begin{align}\label{eq:not-1}
        \sum_{o':o'\succeq a}{x(o')} &<\sum_{o':o'\succeq a}{y(o')}
    \end{align}
    and 
    \begin{align}\label{eq:minimal}
        \sum_{o':o'\succ o}{x(o')} &\geq \sum_{o':o'\succ o}{y(o')}.
    \end{align}
    Define 
    \begin{align}\label{eq:def-h-star}
        h^*=\sum_{o':o\succeq o'}{y(o')}
    \end{align}
    and notice that $h^*<1$, due to Equation (\ref{eq:not-1}).
    Equations (\ref{eq:not-1}), (\ref{eq:minimal}), and (\ref{eq:def-h-star}) now yield
    \begin{align}\label{eq:h-star-defn-lower}
        \sum_{o':o\succeq o'}{x(o')} &= 1-\sum_{o':o'\succeq a}{x(o')}>1-\sum_{o':o'\succeq a}{y(o')}=\sum_{o':o\succeq o'}{y(o')}=h^*
    \end{align}
    and
    \begin{align}\nonumber
        \sum_{o':o\succ o'}{x(o')} &=1-\sum_{o':o'\succeq o}{x(o')}\leq 1-\sum_{o':o'\succeq o}{y(o')}\\\label{eq:h-star-defn-upper}
        &=\sum_{o':o\succ o'}{y(o')}\leq \sum_{o':o\succeq o'}{y(o')}=h^*. 
    \end{align}
    From Equations~(\ref{eq:h-star-defn-lower}) and~(\ref{eq:h-star-defn-upper}), we conclude that $\rep(x,\succ,h^*)=o$. Now, let $b$ be the least preferred option in $O$ that satisfies $b\succ o$ and $y(b)>0$. Such an option exists since Equation (\ref{eq:not-1}) implies that $\sum_{o':o\succeq o'}{y(o')}<1$. Using Equation~(\ref{eq:def-h-star}), we have
    \begin{align}\label{eq:last-1}
        \sum_{o':b\succeq o'}{y(o')} & \geq y(b)+\sum_{o':o\succeq o'}{y(o')}>h^*
    \end{align}
    and
    \begin{align}\label{eq:last-2}
        \sum_{o':b\succ o'}{y(o')} &= \sum_{o':o\succeq o'}{y(o')}=h^*.
    \end{align}
    Equations~(\ref{eq:last-1}) and~(\ref{eq:last-2}) imply $\rep(y,\succ,h^*)=a\succ o=\rep(x,\succ,h^*)$, a contradiction. The ``if'' part of Theorem~\ref{thm:stringent} follows, completing the proof.
\end{proof}

\section{Voting}\label{sec:voting}
We first study the quantile agent utility assumption in a voting setting with a set $A$ of $m$ alternatives and a set $N$ of $n$ agents (voters). Each agent $i\in N$ has a quantile parameter $h_i\in [0,1)$. We use $\h=(h_1, ..., h_n)$ to denote the vector of the quantile parameters. A voting profile $\profile=(\mathord{\succ}_1, \dots, \mathord{\succ}_n)$ consists of the preference $\succ_i$ of each agent; $\succ_i$ is a strict ordering over the alternatives. We use the term {\em voting instance} and identify the above scenario with the tuple $(N,A,\profile,\h)$. We adapt the notion of efficiency for lotteries and agents with quantile utilities as follows.

\begin{definition}[efficiency]\label{defn:eff}
    Given a voting instance $(N,A,\profile,\h)$, a lottery $x$ over the alternatives of $A$ is {\em efficient} if there is no other lottery $y$ such that $\rep(y,\succ_i,h_i)\succeq_i \rep(x,\succ_i,h_i)$ for every agent $i\in N$ and there is an agent $i^*\in N$ such that $\rep(y,\succ_{i^*},h_{i^*})\succ_{i^*} \rep(x,\succ_{i^*},h_{i^*})$.
\end{definition}
Efficiency (as well as all axioms considered in this paper) is studied with respect to a single quantile parameter vector. A definition similar in spirit to efficiency under stochastic dominance (as well as to necessary Pareto optimality; see~\citealp{ABL+19}) would require to consider all quantile parameter vectors. Unfortunately, such definitions are too stringent to be useful, as the next statement illustrates and will not be considered further. 

\begin{restatable}{theorem}{thmvotingAllquantile}\label{thm:votingAllquantile}
Consider a set $N$ of $n$ agents, a set $A$ of alternatives, and a preference profile $\profile$ with the preferences of the agents in $N$ over the alternatives in $A$. There is a lottery that is efficient for the voting instance $(N,A,\profile,\h)$ for every quantile parameter vector $\h\in [0,1)^n$ if and only if all agents have the same alternative ranked first.
\end{restatable}

\begin{proof}
To prove the ``if'' part of the theorem, assume that the same alternative $a\in A$ is in the top position of the preference $\succ_i$ of every agent $i\in N$. Then, the lottery $x$ with $x(a)=1$ and $x(o)=0$ for $o\in A\setminus\{a\}$ makes alternative $a$ the representative of each agent $i$ for every value of her quantile parameter $h_i\in [0,1)$. Clearly, the lottery $x$ is efficient.

For the ``only if'' part, let $L$ be the set of top-ranked alternatives and assume that $|L|\geq 2$. 
First, consider the quantile parameter vector $\h^1=(0,0,...,0)$. We claim that a lottery is efficient for $(N,A,\profile,\h^1)$ only if it returns one of the alternatives with probability $1$. Indeed, assume otherwise and let $S\subseteq A$ be the set of alternatives with positive probability under lottery $x$, i.e., $x(o)>0$ for $o\in S$ and $x(o)=0$ for $o\in A\setminus S$, with $|S|\geq 2$. By definition, the representative $\rep(x,\succ_i,0)$ of agent $i\in N$ is the alternative in $S$ that is ranked lowest among the alternatives in $S$. Then, lottery $x$ is dominated by the lottery $x'$ defined as $x'(a)=1$ for the alternative $a\in S$ that is ranked higher than any other alternative in $S$ by agent $1$ and $x'(o)=0$ for $o\in A\setminus\{a\}$. In this way, agent $1$ has a strictly better representative under $x'$ compared to $x$, while the representative under $x'$ of any other agent is at least as good as the one under $x$.

Next, set $\ell=|L|\geq 2$ and consider the quantile parameter vector $\h^2=(1-1/\ell, 1-1/\ell, ..., 1-1/\ell)$. The lottery $y$ with $y(o)=1/\ell$ for $o\in L$ and $y(o)=0$ for $o\in A\setminus L$ yields the top-ranked alternative of each agent as her representative. Any other lottery $y'$ must have $y'(o)<1/\ell$ for some alternative $o\in L$, implying that the agent who has alternative $o$ as her top choice prefers lottery $y$ to $y'$. Thus, lottery $y$ dominates every other lottery (i.e., it is the unique efficient lottery for the quantile parameter vector $\h^2$), including the set of (deterministic) lotteries that are efficient for the quantile parameter vector $\h^1$. We conclude that, if $|L|\geq 2$, no lottery is efficient for both quantile parameter vectors $\h^1$ and $\h^2$, completing the ``only if'' part of the proof.
\end{proof}

A (randomized) voting rule\footnote{Randomized voting rules are also called {\em social decision schemes} or {\em probabilistic social choice functions} in the literature; e.g., see the survey by~\citet{B17}.} $R$ takes as input a profile and returns a lottery over the alternatives in $A$. Given a profile $\profile$ and an alternative $a\in A$, we denote by $R_a(\profile)$ the probability assigned to alternative $a$ when the voting rule $R$ is applied on the profile $\profile$.  A randomized voting rule is efficient if it produces an efficient lottery. Next, we define strategyproofness for agents with quantile utilities.

\begin{definition}[strategyproofness]\label{defn:sp}
Given a set $N$ of $n$ agents with quantile parameter vector $\h$ and a set $A$ of $m$ alternatives, the voting rule $R$ is {\em strategyproof} for the domain $(N,A,\cdot,\h)$ if for every preference profile $\profile$, agent $i$, and preference $\succ'_i$, it holds
$\rep(R(\profile), \mathord{\succ}_i,h_i)\succeq_i \rep(R(\profile_{-i},\mathord{\succ}'_i), \mathord{\succ}_i,h_i)$.
\end{definition}
\noindent As usual, the notation $(\profile_{-i}, \mathord{\succ}'_i)$ is used to denote the profile that is obtained from $\profile$ when agent $i$ changes her preference to $\succ'_i$.

A natural task is to characterize strategyproof voting rules for agents with quantile utilities. Towards this, we first define monotonic voting rules for voting instances with two alternatives.

\begin{definition}[monotonicity]
    Consider the voting rule $R$ applied on profiles consisting of the preferences of a set $N$ of agents and a set $A=\{a,b\}$ of two alternatives. The rule $R$ is {\em monotonic} if $R_a(\profile) \geq R_a(\profile_{-i},\succ'_i)$ for every preference profile $\profile$, every agent $i\in N$ with $a\succ_i b$, and preference $\succ'_i$ with $b\succ'_i a$. 
\end{definition}

We show the relation between monotonic voting rules and strategyproofness.

\begin{theorem}
    A monotonic voting rule for two alternatives is strategyproof for any quantile parameter vector. For any non-monotonic rule $R$, there exists a quantile parameter vector $\h$ such that $R$ is not strategyproof.  
\end{theorem}

\begin{proof}
Consider the monotonic voting rule $R$ applied on a preference profile $\profile$ with two alternatives $a$ and $b$. For each agent $i\in N$ with $a\succ_i b$, it suffices to consider the case where $\rep(R(\profile),\succ_i,h_i)=b$ and show that $\rep(R(\profile_{-i},\succ'_i),\succ_i,h_i)=b$ as well, where $b\succ'_i a$. Assume otherwise that $\rep(R(\profile_{-i},\succ'_i),\succ_i,h_i)=a$. By the definition of $\rep(R(\profile),\succ_i,h_i)$, we have that $R_b(\profile)>h_i$, i.e., $R_a(\profile)<1-h_i$. By the definition of $\rep(R(\profile_{-i},\succ'_i),\succ_i,h_i)$, we have that $R_b(\profile_{-i},\succ'_i)\leq h_i$, i.e., $R_a(\profile_{-i},\succ'_i) \geq 1-h_i$. Thus, $R_a(\profile)<R_a(\profile_{-i},\succ'_i)$, contradicting the monotonicity of $R$.

Consider a non-monotonic voting rule $R$, a preference profile $\profile$, an agent $i$ with $a\succ_i b$, and another preference $\succ'_i$ with $b\succ'_i a$ such that $t_1=R_a(\profile)<R_a(\profile_{-i},\succ'_i)=t_2$. By setting $h_i=1-\frac{t_1+t_2}{2}$ (notice that this definition yields a valid quantile parameter with $0<h_i<1$), we have $\rep(R(\profile),\succ_i,h_i)=b$ and $\rep(R(\profile_{-i},\succ'_i),\succ_i,h_i)=a$, which violates the strategyproofness of $R$. Indeed, $R_b(\profile)=1-R_a(\profile)=1-t_1>1-\frac{t_1+t_2}{2}=h_i$ (hence, $\rep(R(\profile),\succ_i,h_i)=b$), and $R_b(\profile_{-i},\succ'_i)=1-R_a(\profile_{-i},\succ'_i)=1-t_2\leq 1-\frac{t_1+t_2}{2}=h_i$ and $R_a(\profile_{-i},\succ'_i)=t_2>t_1\geq 0$ (hence, $\rep(R(\profile_{-i},\succ'_i),\succ_i,h_i)=a$).
\end{proof}

When efficiency and strategyproofness are defined using stochastic dominance, the celebrated result of \citet{G77chance} states that random dictatorships are the only strategyproof and efficient\footnote{Efficiency, here, is equivalent to {\em ex post} efficiency. A voting rule is a random dictatorship if it returns the top alternative of a randomly selected agent, according to some probability distribution.} voting rules. In the remainder of this section, we explore whether analogous characterizations for strategyproof and efficient rules exist in our model (under Definitions~\ref{defn:eff} and~\ref{defn:sp}).

For two alternatives, we define the randomized voting rule $Q$ (and call it $Q$-{\em plurality}) as follows. Consider the voting instance $(N,\{a,b\},\profile,\h)$ with two alternatives $\{a,b\}$ and, without loss of generality, assume that alternative $s$ is the plurality winner, i.e., the alternative that is preferred by the majority\footnote{We assume that ties are always broken in favor of the same alternative.} of the agents in $\profile$. $Q$-plurality returns the plurality winner $a$ as winning alternative with probability $Q_a(\profile)=1-\min_{i:a\succ_i b}{h_i}$ (and returns alternative $b$ with probability $Q_b(\profile)=\min_{i:a\succ_i b}{h_i}$).

\begin{theorem}
$Q$-plurality is an efficient and strategyproof rule for voting instances with two alternatives.
\end{theorem}

\begin{proof}
Consider a voting instance $(N,\{a,b,\},\profile,\h)$ in which (without loss of generality) alternative $a$ is the plurality winner. Denote by $W=\{i\in N:a\succ_i b\}$ the set of agents who prefer alternative $a$ to alternative $b$. Notice that $\rep(Q(\profile),\succ_j,h_j)=a$ for every agent $j\in W$; indeed, $Q_b(\profile)=\min_{i\in W}{h_i}\leq h_j$ and $Q_a(\profile)+Q_b(\profile)=1>h_j$ (by definition). Let $L=\{i\in N: b\succ_i a, \rep(Q(\profile),\mathord{\succ}_i,h_i)=a\}$ be the set of agents who have alternative $a$ as their least preferred alternative and as representative. We will prove the efficiency of $Q$-plurality by showing that any lottery $x$ that makes alternative $b$ the representative for some agent $i^*$ in $L$ (and is, thus, more preferable than $Q$ for this agent) makes the least preferred alternative the representative for some agent in $W$ (and is, thus, less preferable than $Q$ for this agent).

Consider the lottery $x$ that satisfies $
\rep(x,\succ_j,h_j)=b$ for some agent $j\in L$ and let $i^*$ be an agent in $W$ with minimum quantile parameter, i.e., $h_{i^*}=\min_{i\in W}{h_i}$. Since $\rep(x,\succ_j,h_j)=b$, $\rep(Q(\profile),\succ_j,h_j)=a$, and by the definition of lottery $Q$ and the quantile parameter $h_{i^*}$, we have
\begin{align*}
    x(a) &\leq h_j <Q_a(\profile)=1-\min_{i\in W}{h_i}=1-h_{i^*}.
\end{align*}
Equivalently, $x(b)=1-x(a)>h_{i^*}$, which implies that $\rep(x,\succ_{i^*},h_{i^*})=b$, as desired.

To prove strategyproofness, it suffices to consider an agent $j\in L$ and show that alternative $a$ is still her representative under $Q$-plurality after she changes her preference to $a\succ'_j b$. Clearly, alternative $a$ is still the plurality winner in profile $(\profile_{-j},\succ'_j)$ and, hence,
\begin{align*}
    Q_a(\profile_{-j},\succ'_j) &= 1-\min_{i\in W\cup\{j\}}{h_i} \geq 1-\min_{i\in W}{h_i}=Q_a(\profile) >h_j,
\end{align*}
which implies that $\rep(Q(\profile_{-j},\succ'_j),\succ_j,h_j)=a$, as desired. The last inequality follows since $j\in L$ and, hence, $b\succ_j a$ and $\rep(Q(\profile),\succ_j,h_j)=a$.
\end{proof}

For more than two alternatives, the picture changes. Our next statement demonstrates that, rather unsurprisingly, deterministic dictatorships are the only efficient and strategyproof voting rules when the agents have small quantile parameters. However, positive results are possible for quantile parameters higher than $1/3$. Some of them are rather surprising, e.g., the statements in case (b).

\begin{restatable}{theorem}{thmvoting}\label{thm:voting}
Consider a set $N$ of $n$ agents with the quantile parameter vector $\h$ and a set of alternatives $A=\{a,b,c\}$. 
\begin{enumerate}
    \item[(a)] If $\h\in [0,1/3)^n$, any efficient and strategyproof voting rule is a deterministic dictatorship. 
    \item[(b)] If $\h\in [1/3, 1/2)^n$, and $n > 2$ there exists an efficient and strategyproof rule different from deterministic dictatorships. No such rule exists if $n = 2$ and $\h=(h,h)$ with $h\in [1/3,1/2)$.
    \item[(c)] If $\h\in [1/2,2/3)^n$, the voting rule that returns each of the two alternatives with the highest plurality score with probability $1/2$ is efficient and strategyproof. 
    \item[(d)] If $\h\in [2/3,1)^n$, the voting rule that returns each alternative with probability $1/3$ is efficient and strategyproof.
\end{enumerate}
\end{restatable}

\begin{proof}
{\bf (a)} Consider any lottery $x$ over the alternatives in $A$ assume that $x(a)=\max_{o\in A}{x(o)}$. Then, $x(a)\geq 1/3$ and $\sum_{o:a\succeq_i o}{x(o)}\geq 1/3>h_i$ for every $i\in N$. This implies that $a\succeq_i\rep(x,\succ_i,h)$, meaning that the lottery $x$ is either dominated by the deterministic lottery $y(a)=1$ (and $y(b)=y(c)=0$) or the two lotteries are equivalent in the sense that, for any agent, they yield the same alternative as representative. Hence, any efficient voting rule can be considered as deterministic in this case, and the statement follows from the Gibbard-Satterthwaite theorem~\citep{Gib73,Sat75}.

\paragraph{\bf (b)}
Before proving the statements, we argue that we can assume that there are only four types of efficient lotteries, namely the uniform lottery, and the three lotteries that put probability $1$ on a single alternative. All other efficient lotteries return the same representatives as one of those four.
To prove this claim, consider any lottery $p$.
If there is no alternative $a$ such that $p(a) > \frac{1}{2}$, then no agent gets their top choice. Furthermore, observe that the uniform lottery guarantees each agent their second-best choice. Thus, we can assume that whenever there is no alternative with probability larger than $\frac{1}{2}$, an efficient voting rule returns the uniform lottery (since both $p$ and the uniform lottery return the same representatives due to $p$ being efficient).
So, consider the case in which there exists an alternative $a$ such that $p(a) > \frac{1}{2}$. We claim that $p$ returns the same representatives as a lottery $q$ that puts probability $1$ on alternative $a$. This follows since every agent has a quantile parameter strictly larger than $\frac{1}{2}$ and, thus, cannot have a representative that is preferred to $a$ (since the probability of all other alternatives is at most $\frac{1}{2}$). Therefore, in order to be efficient, $p$ returns the same representative as $q$ for every agent.

In the following, we give a voting rule that is strategyproof and efficient:
If there exists a strict majority winner, then it gets chosen with probability $1$ (Case 1). If an alternative is the top choice of precisely half the agents, it gets selected if and only if it is the top choice of a special agent $v_s$ (Case 2). If the first two cases do not occur and there exists an alternative that is never the last choice then this alternative gets selected (Case 3). Finally, if none of the cases above applies, the voting rule returns the uniform lottery (Case 4).

Observe that the rule is well-defined for Case 3 since if both Case 1 and Case 2 did not apply, it holds that every alternative is the top choice of at least one agent. Therefore, by the pigeonhole principle, at most one alternative can exist that is never the last choice.

First, we want to argue why the voting rule is efficient. Observe that placing probability $1$ on an alternative that is the top choice of at least one agent is always efficient (Case 1 and 2). Furthermore, an alternative that is never the last choice is efficient as long as no other alternative is always the top choice (Case 3).
Finally, if every alternative is the last choice of at least one agent then the uniform lottery is efficient (Case 4).

Next, we argue why the voting rule is strategyproof.
So, without loss of generality, let $a \succ b \succ c$ be the preference of any agent $i$. A manipulation is only possible if the voting rule either returns $b, c$, or the uniform lottery. In the following, we do a case distinction over which case of the rule applies:
If the Case 1 or 2 applies, then either a strict majority has $b$ or $c$, respectively, as their top choice, or a weak majority including the special agent $v_s$, who then has to be different to $i$. Therefore, regardless of which preferences agent $i$ reports, still $b$ or $c$ get returned.
So assume that Case 3 applies. This directly implies that alternative $b$ is returned (since $c$ is the last choice of agent $i$). Furthermore, since Cases 1 and 2 did not apply, every alternative is the top choice at least once. Therefore, by the pigeonhole principle, alternative $a$ must be the last choice for at least one agent. So, Case 3 cannot return alternative $a$, no matter what preferences agent $i$ reports. Thus, in order for $a$ to be returned, the number of agents submitting $a$ as their top choice has to strictly increase, which does not happen, regardless of the preferences that agent $i$ reports.
It remains to show that no manipulation is possible for Case 4. Since agent $i$'s representative for the uniform lottery is $b$, a successful manipulation means that $a$ is returned. Again, there exists at least one other agent having $a$ so there does not exist any manipulation of agent $i$ such that alternative $a$ gets returned due to Case 3. But, as above, the number of agents approving $a$ has to increase so that alternative $a$ gets returned for the Cases 1 and 2. 

Observe that for $n = 2$ the described rule is a deterministic dictatorship with $v_s$ being the dictator.
We now prove that this is necessary. Let $\h=(h_1,h_2)$ with $h_1,h_2\in [1/3,1.2)$. By the discussion in the first paragraph of part ({\bf b}), we can restrict our attention to voting rules that return either some alternative deterministically or the uniforma lottery in each profile. Clearly, any voting rule that never returns the uniform lottery is a dictatorship according to the Gibbard-Satterthwaite theorem~\citep{Gib73,Sat75}. We will complete the proof of the characterization by showing that each efficient voting rule for two agents has this property.

Assume otherwise, and let $R$ be an efficient voting rule that returns the uniform lottery in some profile. This means that for both agents, their representative is ranked second. We can verify that the only case in which such a lottery is efficient is for a ``palindromic'' profile in which the two agents have different top-ranked alternatives and the same alternative ranked second, which is the representative of both. Any other case violates efficiency since the lottery would be dominated by one which returns some top-ranked alternative with probability $1$.

Without loss of generality, we can assume that in the palindromic ranking, agent $1$ has the ``left'' preference $a\succ^L c \succ^L b$ and agent $2$ has the ``up'' preference 
$b\succ^U c \succ^U a$. We also define the ``right'' preference $a\succ^R b \succ^R c$ and the ``down'' preference $b\succ^D a \succ^D c$ and the profiles $\profile^{LU}$, $\profile^{RU}$ and $\profile^{RD}$, depending on whether agent $1$ has the left (L) or the right (R) preference and agent $2$ has the up (U) or down (D) preference. The four profiles are depicted in Figure~\ref{fig:profiles}.

\begin{figure}[ht]
    \centering
    \begin{minipage}{0.45\linewidth}
        \centering
        \begin{tabular}{rl}
            & profile $\profile^{LU}$ \\\hline
            $1$: & $a\succ^L c\succ^L b$\\ 
            $2$: & $b\succ^U c\succ^U a$ 
        \end{tabular}
    \end{minipage}
    \begin{minipage}{0.45\linewidth}
         \centering
        \begin{tabular}{rl}
            & profile $\profile^{RU}$ \\\hline
            $1$: & $a\succ^R b\succ^R c$\\ 
            $2$: & $b\succ^U c\succ^U a$ 
        \end{tabular}
    \end{minipage}

    \vspace{1em} 

    \begin{minipage}{0.45\linewidth}
        \centering
        \begin{tabular}{rl}
            & profile $\profile^{LD}$ \\\hline
            $1$: & $a\succ^L c\succ^L b$\\ 
            $2$: & $b\succ^D a\succ^D c$ 
        \end{tabular}
    \end{minipage}
    \begin{minipage}{0.45\linewidth}
         \centering
        \begin{tabular}{rl}
            & profile $\profile^{RD}$ \\\hline
            $1$: & $a\succ^R b\succ^R c$\\ 
            $2$: & $b\succ^D a\succ^D c$ 
        \end{tabular}
    \end{minipage}
    \caption{The four profiles used in the proof of Theorem~\ref{thm:voting}b, second part.}
    \label{fig:profiles}
\end{figure}

In the non-palindromic profiles $\profile^{RU}$, $\profile^{LD}$, and $\profile^{RD}$, some of the alternatives $a$ and $b$ is returned with probability $1$ by $R$. Hence, the same alternative $a$ or $b$ is the representative of both agents. First, assume that alternative $a$ is the representative of both agents in profile $\profile^{RD}$, i.e., $\rep(R(\profile^{RD}),\succ^R,h_1)=\rep(R(\profile^{RD}),\succ^D,h_2)=a$. Then, alternative $a$ should also be the representative of both agents in profile $\profile^{RU}$, i.e., $\rep(R(\profile^{RU}),\succ^R,h_1)=\rep(R(\profile^{RU}),\succ^U,h_2)=a$. Indeed, if alternative $b$ was the representative of agent $2$ in profile $\profile^{RU}$, then $\rep(R(\profile^{RU}),\succ^D,h_2)=b\succ^D a=\rep(R(\profile^{RD}),\succ^D,h_2)$ and agent $2$ could manipulate the voting rule $R$ at profile $\profile^{RD}$ by misreporting the preference $\succ^U$ instead of $\succ^D$. 

Recall that, in the palindromic profile $\profile^{LU}$, agent $1$ has the second-ranked alternative $c$ as representative. Hence, $\rep(R(\profile^{RU}),\succ^L,h_1)=a \succ^L c=\rep(R(\profile^{LU}),\succ^L,h_1)$, and agent $1$ could manipulate the voting rule $R$ at profile $\profile^{LU}$ by misreporting the preference $\succ^R$ instead of $\succ^L$. This contradicts our assumption that the voting rule $R$ returns the uniform lottery in the palindromic profile $\profile^{LU}$. 

Now, assume that alternative $b$ is the representative of both agents in profile $\profile^{RD}$, i.e., $\rep(R(\profile^{RD}),\succ^R,h_1)=\rep(R(\profile^{RD}),\succ^D,h_2)=b$. Then, alternative $b$ should also be the representative of both agents in profile $\profile^{LD}$, i.e., $\rep(R(\profile^{LD}),\succ^L,h_1)=\rep(R(\profile^{LD}),\succ^D,h_2)=b$. Indeed, if alternative $a$ was the representative of agent $1$ in profile $\profile^{LD}$, then $\rep(R(\profile^{LD}),\succ^R,h_1)=a\succ^R b=\rep(R(\profile^{RD}),\succ^R,h_1)$ and agent $1$ could manipulate the voting rule $R$ at profile $\profile^{RD}$ by misreporting the preference $\succ^L$ instead of $\succ^R$. 

Since the voting rule $R$ makes alternative $c$ the representative of agent $2$ in the palindromic profile $\profile^{LU}$, we have $\rep(R(\profile^{LD}),\succ^U,h_2)=b \succ^U c=\rep(R(\profile^{LU}),\succ^U,h_2)$, and agent $2$ could manipulate the voting rule $R$ at profile $\profile^{LU}$ by misreporting the preference $\succ^D$ instead of $\succ^U$. This again contradicts our assumption that the voting rule $R$ returns the uniform lottery in the palindromic profile $\profile^{LU}$. 

\paragraph{\bf (c)} Consider a profile $\profile$ and let $a$ and $b$ be the two alternatives with the highest plurality score. Then, the lottery $x$ with $x(a)=x(b)=1/2$ and $x(c)=0$ yields as representative in each agent her highest-ranked alternative among $a$ and $b$. 

We first show that $x$ is efficient. Notice that lottery $x$ can be dominated only by some lottery $y$ that makes alternative $c$ the representative of an agent $j$ who has alternative $c$ ranked first. By the definition of the representative $
\rep(y,\succ_j,h_j)$ and the assumption on $h_j$, it must be $y(a)+y(b)\leq h_j<2/3$. Thus, lottery $y$ must assign probability smaller than $1/3$ to at least one of the alternatives $a$ and $b$. Assume, without loss of generality, that $y(a)<1/3$ and consider an agent $i$ with alternative $a$ as her top choice (such an agent exists since both alternatives $a$ and $b$ have at least as high plurality score as alternative $c$). Then, $y(b)+y(c)=1-y(c)>2/3>h_i$, meaning that $a\succ_i\rep(y,\succ_i,h_i)$. Hence, lottery $y$ does not dominate lottery $x$ and lottery $x$ is efficient.

To prove strategyproofness, it suffices to consider an agent $j\in N$ with $c$ as top-ranked alternative and, thus, her second-ranked alternative as representative under lottery $x$. Now, observe that any deviating preference $\succ'_j$ by the agent $j\in N$ cannot increase the plurality score of $c$ or decrease the plurality score of $a$ and $b$ in $(\profile_{-j},\succ'_j)$ compared to $\profile$. Thus, $a$ and $b$ are still the two alternatives with the highest plurality score, and the lottery returned by the rule is the same.

\paragraph{\bf (d)} In this case, the uniform lottery with $x(a)=x(b)=x(c)=1/3$ makes the top-ranked alternative of each agent her representative. Hence, this rule is trivially efficient and strategyproof.
\end{proof}

With our Theorem~\ref{thm:voting}, we have made some initial progress towards answering the following challenging open problem.
\begin{question}
Characterize the efficient and strategyproof randomized voting rules for voting settings with at least three alternatives, any number of agents, and all quantile parameter vectors.
\end{question}

\section{One-sided matching}\label{sec:matching}
In this section, we consider settings with a set $N$ of $n$ agents and a set $M$ of $n$ items. Each agent $i \in N$ has a quantile parameter $h_i\in [0,1)$ and a preference order $\succ_i$ over the items in $M$. A matching mechanism takes as input a preference profile $\profile = (\succ_1, ...,\succ_n)$ and the vector $\h$ of quantile parameters, and returns a lottery over perfect matchings between the sets $N$ and $M$. Thus, a lottery $\x$ assigns  probability $x_{ig}$ for agent $i \in N$ and item $g \in M$ such that $\sum_{i\in N} x_{ig} = 1$ for each $g \in M$ and $\sum_{g\in M} x_{ig}=1$ for each $i \in N$. We usually view the lottery $\x$ over matchings as $n$ lotteries over the items in $M$, with one lottery $x_i$ for each agent $i\in N$.

\subsection{Social welfare maximization, efficiency, and strategyproofness}
We begin our study of lotteries over matchings by focusing on the maximization of (variants of) social welfare. The simplest definition of social welfare in our one-sided matching setting is the number of agents who have their top preferred item as representative. Maximizing it turns out to be an easy goal.
\begin{restatable}{theorem}{thmonesidesw}\label{thm:one-sided-sw}
    Given a one-sided matching instance $(N,M,\profile,\h)$, a lottery that maximizes the number of agents having their top-ranked item as representative can be computed in polynomial time.
\end{restatable}

\begin{proof}
For each item $g\in M$, denote by $S_g$ the set of agents having item $g$ as their top choice.  An agent from $S_g$ can have item $g$ as representative only if the lottery assigns her to item $g$ with probability at least $1-h_i$. Denote by $\overline{S}_g$ a subset of $S_g$ of maximum cardinality that satisfies $\sum_{i\in \overline{S}_g}{(1-h_i)}\leq 1$; this condition is necessary so that there is a lottery that makes item $g$ representative for all agents in $\overline{S}_g$. Then, the maximum number of agents having their top item as representative is at most $\sum_{g\in M}{|\overline{S}_g|}$.

We now present a lottery under which a maximum number of $\sum_{g\in M}{|\overline{S}_g|}$ agents have their top item as representative. Notice that for every two different items $g_1,g_2\in M$, the sets $\overline{S}_{g_1}$ and $\overline{S}_{g_2}$ are disjoint. Thus, by setting $x_{ig}=1-h_i$ for every $g\in M$ and $i\in \overline{S}_g$, we get that all agents in $\cup_{g\in M}{\overline{S}_g}$ have their top item as representative. By the definition of set $\overline{S}_g$, we have the condition $\sum_{i\in \overline{S}_g}{x_{ig}}\leq 1$ for every item $g$. Hence, we can trivially complete $\x$ and get a valid lottery by setting the values $x_{ig}$ for every other agent-item pair $(i,g)$ with $i\not\in \overline{S}_g$ so that $\sum_{i\in N}{x_{ig}}=1$ for every item $g\in M$ and $\sum_{g\in M}{x_{ig}}=1$ for every agent $i\in N$.

Notice that, for a given item $g\in M$, the set $\overline{S}_g$ can be easily computed by starting with the empty set, considering the agents in set $S_g$ in monotone non-increasing order of their quantile parameter, and including an agent in $\overline{S}_g$ as long as the sum of the quantities $1-h_i$ in $\overline{S}_g$ does not exceed $1$.
\end{proof}

Interestingly, the problem becomes NP-hard if we aim to maximize the number of agents having some of their {\em two} top-ranked items as representative. The proof of the next statement appears in Appendix~\ref{app:sec:omitted-proofs}. The notation $\rank_i(a)$ is used to denote the position of item $a$ in the preference $\succ_i$ of agent $i$.

\begin{restatable}{theorem}{thmonesidenp}\label{thm:one-sided-NP}
    Given a one-sided matching instance $(N, M, \mathord{\profile}, \h)$, an integer rank requirement $r_i\in \{1,2\}$ for every agent $i \in N$, and a non-negative integer $k$, deciding whether there exists a lottery $\x$ such that $|\left\{i\in N: \rank_i(\rep(x_i,\succ_i,h_i))\in [r_i]\right\}| \geq k$ is NP-hard. \end{restatable}

We now adapt the definitions of efficiency and strategyproofness for lotteries over matchings and matching mechanisms. 
\begin{definition}[efficiency for lotteries over matchings]
    Given a one-sided matching instance $(N,M,\profile,\h)$, a lottery $\x$ over matchings is {\em efficient} if there is no other lottery $\y$ such that $\rep(y_i,\mathord{\succ}_i,h_i)\succeq_i \rep(x_i,\succ_i,h_i)$ for every agent $i\in N$ and there is an agent $i^*\in N$ such that $\rep(y_{i^*},\succ_{i^*},h_{i^*})\succ_{i^*} \rep(x_{i^*},\succ_{i^*},h_{i^*})$.
\end{definition}

\begin{definition}[strategyproofness of matching mechanisms]
Given a set $N$ of $n$ agents with quantile parameter vector $\h$ and a set $M$ of $n$ items, the matching mechanism $R$ is {\em strategyproof} for the domain $(N,M,\cdot,\h)$ if for every preference profile $\profile$, agent $i\in N$, and preference $\succ'_i$, it holds
$\rep(R(\profile), \mathord{\succ}_i,h_i)\succeq_i \rep(R(\profile_{-i},\succ'_i), \mathord{\succ}_i,h_i)$.
\end{definition}

We will prove that efficiency and strategyproofness are compatible for agents with quantile utilities. The main component of our proof is a simple set of linear inequalities that check whether lotteries over matchings that satisfy given requirements for the ranks of the representatives of the agents exist. For a given one-sided matching instance $(N,M,\profile,\h)$ and a vector of rank requirements $\rr=(r_1, r_2, ..., r_n)$ with integer $r_i\in [n]$, the following linear program is feasible if and only if there is a lottery over matchings that satisfy the rank requirement $r_i$ for the representative of each agent $i\in N$.
\begin{align*}
    \sum_{g:\rank_i(g)\leq r_i}{x_{ig}} \geq 1-h_i,& \quad \forall i\in N\\
    \sum_{g\in M}{x_{ig}}=1, &\quad \forall i\in N\\
    \sum_{i\in N}{x_{ig}}=1, &\quad \forall g\in M
\end{align*}
We will refer to this linear program as $\LP(\profile;\rr)$.

The compatibility of efficiency and strategyproofness can be obtained via a variant of the classical {\em serial dictatorship} mechanism for computing integral matchings. That mechanism considers the agents one by one according to a predefined order. Whenever an agent is considered, she is assigned her most preferred item that has not been assigned to any agent before. Our extension to lotteries over matchings and agents with quantile utilities below employs linear programming to (solve the linear programs $\LP(\profile;\rr)$ and) guarantee the most preferred representative for each agent considered.

\paragraph{Serial dictatorship (SD) mechanism:} On input a one-sided matching instance $(N,M,\profile,\h)$ with $n$ agents/items, mechanism SD starts with a vector $\rr=(n, ..., n)$ of minimum rank requirements for all agents and considers the agents one by one in increasing order of their ids. When the agent $i\in N$ is considered, the mechanism computes the minimum rank $t$ for agent $i$ so that the linear program $\LP(\profile;\rr_{-i},t)$ is feasible and updates $\rr$ by setting $r_i=t$. After all agents have been considered, an arbitrary lottery over matchings that satisfies $\LP(\profile;\rr)$ is returned as the output of the mechanism. 

We have the following statement, which we will improve considerably until the end of this section.

\begin{restatable}{theorem}{thmonesideeffsp}\label{thm:one-sided:eff+sp}
Mechanism SD is efficient and strategyproof.
\end{restatable}

\begin{proof}
The crucial property of mechanism SD is that the final linear program $\LP(\profile;\rr)$ has as solutions all those lotteries which make representative for agent $1$ her top-ranked item and representative for agent $i\geq 2$ her highest-ranked item under rank-constraints for the representative items for agents $1, ..., i-1$. 

To prove efficiency, let $\x$ be the lottery returned by the mechanism as the solution of the linear program $\LP(\profile;\rr)$. For the sake of contradiction, assume that there is another lottery $\y$ such that $\rank_i(y_i,\succ_i,h_i)\leq \rank_i(x_i,\succ_i,h_i)$ for every agent $i\in N$ and $\rank_{i^*}(y_{i^*},\succ_{i^*},h_{i^*})<\rank_{i^*}(x_{i^*},\succ_{i^*},h_{i^*})$ for some agent $i^*\in N$. Let $i'\in N$ be the agent of minimum id with $r'_{i'}=\rank_{i'}(y_{i'},\succ_{i'},h_{i'})< r_i$, i.e., $\rank_{i}(y_{i},\succ_{i},h_{i})= r_i$ for $i=1, ..., i'-1$. This means that the linear program $\LP(\profile;(r_1, ..., r_{i'-1}, r'_{i'},n, ..., n))$ has lottery $\y$ as solution, contradicting the definition of $r_{i'}$ at round $i'$ of the mechanism.

Strategyproofness follows since when it considers agent $i\in N$, the mechanism restricts the candidate output lotteries to those that make the best possible item a representative for agent $i\in N$, given decisions about the representatives of agents $1, ..., i-1$ in previous rounds. Hence, unilateral misreporting by agent $i$ cannot result to a final lottery that improves her representative item further.
\end{proof}

We now make an observation that stands in sharp contrast to what we know about integral matchings. If we restrict our attention to such matchings where each agent is assigned a distinct item and define efficiency taking only such matchings into account (and ignore lotteries), a characterization result by~\citet{AS98} states that a matching is efficient if and only if it can be produced by the serial dictatorship mechanism. Translated to our context, an efficient lottery over matchings for agents with quantile parameters equal to $0$ can be produced by serial dictatorship when considering the agents in an appropriate order. This is not the case for agents with positive quantile parameters as the next statement shows.

\begin{restatable}{theorem}{thmnosdeffequivalence}\label{thm:no-sd-eff-equivalence}
    There exists a one-sided matching instance $(N,M,\profile,\h)$ and an efficient lottery that cannot be produced by mechanism SD.
\end{restatable}

\begin{proof} 
Consider the following instance with five agents and quantile parameters $h_1 = h_2 = 0.2$, $h_3 = 0.4$, $h_4 = h_5 = 0.6$. Agents $1$, $2$, and $3$ have items $a$, $b$, $c$ as their top preference, respectively. Agent $4$ has items $c$ and $a$ as their first and second preference, respectively. Agent $5$ has items $c$ and $b$ as their first and second preference, respectively. An example of the profile is depicted in the next table; the part of the profile that is not defined above does not affect the argument in the proof.

\begin{table}[h]
\centering
\begin{tabular}{c|c|ccccccccc}
agent & quantile &\multicolumn{9}{c}{preference order}\\\hline
$1$ & $0.2$ & $a$ & $\mathcolor{gray!70}{\succ_1}$ & $\mathcolor{gray!70}{d}$ & $\mathcolor{gray!70}{\succ_1}$ & $\mathcolor{gray!70}{e}$ & $\mathcolor{gray!70}{\succ_1}$ & $\mathcolor{gray!70}{b}$ & $\mathcolor{gray!70}{\succ_1}$ & $\mathcolor{gray!70}{c}$ \\
$2$ & $0.2$ & $b$ & $\mathcolor{gray!70}{\succ_2}$ & $\mathcolor{gray!70}{d}$ & $\mathcolor{gray!70}{\succ_2}$ & $\mathcolor{gray!70}{e}$ & $\mathcolor{gray!70}{\succ_2}$ & $\mathcolor{gray!70}{a}$ & $\mathcolor{gray!70}{\succ_2}$ & $\mathcolor{gray!70}{c}$ \\
$3$ & $0.4$ & $c$ & $\mathcolor{gray!70}{\succ_3}$ & $\mathcolor{gray!70}{d}$ & $\mathcolor{gray!70}{\succ_3}$ & $\mathcolor{gray!70}{e}$ & $\mathcolor{gray!70}{\succ_3}$ & $\mathcolor{gray!70}{a}$ & $\mathcolor{gray!70}{\succ_3}$ & $\mathcolor{gray!70}{b}$ \\
$4$ & $0.6$ & $c$ & $\succ_4$ & $a$ & $\mathcolor{gray!70}{\succ_4}$ & $\mathcolor{gray!70}{d}$ & $\mathcolor{gray!70}{\succ_4}$ & $\mathcolor{gray!70}{e}$ & $\mathcolor{gray!70}{\succ_4}$ & $\mathcolor{gray!70}{b}$ \\
$5$ & $0.6$ & $c$ & $\succ_5$ & $b$ & $\mathcolor{gray!70}{\succ_5}$ & $\mathcolor{gray!70}{d}$ & $\mathcolor{gray!70}{\succ_5}$ & $\mathcolor{gray!70}{e}$ & $\mathcolor{gray!70}{\succ_5}$ & $\mathcolor{gray!70}{a}$ \\
\end{tabular}
\caption{The one-sided matching instance for the proof of Theorem~\ref{thm:no-sd-eff-equivalence}. The lightly colored part of the preference orders is only indicative and does not affect the argument in the proof, which is still valid for every redistribution of the items in those positions of the preferences. }
\label{tab:instance-no-sd-eff-equivalence}
\end{table}
Now, consider the lottery that is defined in Table~\ref{tab:lottery-no-sd-eff-equivalence}. By examining it carefully and using the definitions in Table~\ref{tab:instance-no-sd-eff-equivalence}, we can see that the representatives of agents $1$, $2$, $3$, $4$, and $5$ are $a$, $b$, $c$, $a$, and $b$, respectively.

\begin{table}[h]
\centering
\begin{tabular}{c|cccccc}
& $a$ & $b$ & $c$ & $d$ & $e$ \\\hline
$1$ & $0.8$ & $0$ & $0$ & $0.1$ & $0.1$ \\
$2$ & $0$ & $0.8$ & $0$ & $0.1$ & $0.1$ \\
$3$ & $0$ & $0$ & $0.6$ & $0.2$ & $0.2$ \\
$4$ & $0.2$ & $0$ & $0.2$ & $0.3$ & $0.3$ \\
$5$ & $0$ & $0.2$ & $0.2$ & $0.3$ & $0.3$ \\
\end{tabular}
\caption{The lottery that makes items $a$, $b$, $c$, $a$, and $b$ representatives of agents $1$, $2$, $3$, $4$, and $5$, respectively. }
\label{tab:lottery-no-sd-eff-equivalence}
\end{table}

Now, we argue that this lottery is efficient. Assume it is not. Then, it must be dominated by either a lottery that makes items $a$, $b$, and $c$ representatives of agents $1$, $2$, and $3$, respectively, and either agent $4$ or agent $5$ has item $c$ as representative. Due to symmetry, let us assume that agent $4$ has item $c$ as representative. Then, agent $5$ must have either item $b$ or item $c$ as representative. Since agents $3$ and $4$ with quantile parameters $0.4$ and $0.6$ have their top item $c$ as representative, they must get item $c$ with probabilities $0.6$ and $0.4$. Hence, agent $5$ does not get item $c$, and item $b$ needs probability at least $0.4$ in order to become her representative. But then, agent $2$ with quantile parameter $0.2$ gets her top item $b$ with probability at most $0.6$, contradicting the assumption that it is her representative.

We now argue that serial dictatorship cannot compute a lottery that makes items $a$, $b$, $c$, $a$, and $b$ representatives of agents $1$, $2$, $3$, $4$, and $5$, respectively. Notice that when agents $1$ and $2$ are considered, the SD mechanism does not restrict the probabilities the agent get item $c$. These depend only on agents $3$, $4$, and $5$. If both agents $4$ and $5$ are considered before agent $3$, they will both be assigned item $c$ with probability at least $0.4$ and have it as representative. So, agent $3$ should be considered before at least one of the agents $4$ and $5$. Assume that it is considered before agent $5$. After both agents $3$ and $4$ have been considered, they will have gotten item $c$ with probability at least $0.6$ and $0.4$ and have it as representative. Overall, there is no possibility of computing the lottery above with mechanism SD.
\end{proof}

\subsection{Fairness axioms}
In the following, we define two fairness properties for lotteries over matchings and study their interplay with efficiency and strategyproofness. We begin with the definition of {\em proportionality}.

\begin{definition}[proportionality]\label{defn:prop}
    A lottery $\x$ for the one-sided matching instance $(N,M,\profile,\h)$ is {\em proportional} if $\rank_i(\rep(x_i,\succ_i,h_i))\leq \lceil n(1-h_i)\rceil$ for every agent $i\in N$.
\end{definition}

Notice that $\left\lceil n(1-h_i)\right\rceil$ is exactly the position of the representative of agent $i$ in her preference when the {\em uniform} lottery is used. Thus, a lottery is proportional if each agent weakly prefers it to the uniform lottery. Clearly, the uniform lottery is proportional. 

The first question that arises is whether proportionality is compatible with efficiency and strategyproofness. We prove that this is indeed the case, by running mechanism SD with the agents considered in a non-increasing order in terms of their quantile parameter. We will refer to this mechanism as SDP (serial dictatorship for proportionality).

\begin{theorem}\label{thm:one-sided-eff+prop}
Mechanism SDP is efficient, strategyproof and proportional.
\end{theorem}

\begin{proof}
We use $p_i$ as a shorthand of the rank requirement for agent $i\in N$ from Definition~\ref{defn:prop}, i.e., $p_i=\lceil n(1-h_i)\rceil$. Without loss of generality, assume that the agents' quantile parameters are non-increasing in terms of their ids. The efficiency and strategyproofness of mechanism SDP follows by Theorem~\ref{thm:one-sided:eff+sp}. We complete the proof by proving proportionality. In particular, we will prove that, at the end of the execution of mechanism SDP, the vector $\rr$ satisfies $r_i\leq p_i$, for every $i\in N$.

We use induction on the agents' id. Notice that when agent $1$ is considered, it sets $r_1=1$; clearly, $r_1\leq p_1$. At the step in which agent $i\geq 2$ is considered, let us assume that $r_j\leq p_j$ for $j=1, ..., i-1$. Then, $\LP(\profile;\rr)$ has the additional constraints that reserve probability at least $1-h_j$ for the items that are ranked in the top $r_j$ positions of the preference of agent $j$, for $j=1, ..., i-1$. Hence, the total probability reserved for these items is $\sum_{j=1}^{i-1}{(1-h_j)}$. Thus, assuming that $\LP(\profile,\rr)$ is feasible before considering agent $i$, there is still a total probability of \begin{align*}
    p_i-\sum_{j=1}^{i-1}{(1-h_j)}
    &=\lceil n (1-h_k)\rceil - \sum_{j=1}^{i-1}{(1-h_j)}\\
    &\geq n\cdot (1-h_i)-(i-1)\cdot (1-h_i) \geq 1-h_i 
\end{align*}
to be put on the items ranked in the top $p_i$ positions in the preference of agent $i$ so that $\LP(\profile;\rr_{-i},p_i)$ is feasible. Hence, mechanism SD will set $r_i\leq p_i$ in the current step. 
\end{proof}

Notice that the problem of computing a proportional lottery can be thought of as a special case of the following analogue of the {\em house allocation} problem~\citep{SS74} for randomized outcomes. We are given a matching instance and an {\em endowment lottery} $\e$ and we wish to compute another lottery $\x$ that weakly dominates $\e$, i.e., $\rep(x_i,\succ_i,h_i)\succeq_i \rep(e_i,\succ_i,h_i)$ for every agent $i\in N$. We have proved that SDP is a strategyproof and efficient mechanism for this problem when the endowment lottery is the uniform one. This immediately suggests the next important open question.

\begin{question}
    Is there a strategyproof and efficient mechanism which, given a matching instance and an endowement lottery, computes a lottery that weakly dominates the endowment?
\end{question}

We now define {\em envy-freeness}, our second fairness property.
\begin{definition}[envy-freeness]
    A lottery $x$ for the one-sided matching instance $(N,M,\profile,\h)$ is {\em envy-free} if $\rep(x_i,\succ_i,h_i)\succeq_i \rep(x_j,\succ_i,h_i)$ for every pair of agents $i,j\in N$.
\end{definition}

The definition of envy-freeness is a very natural one. The next statement indicates that, similarly to the relation of these concepts in the literature on fair division, our definitions are such that envy-freeness implies proportionality.

\begin{theorem}
    Envy-freeness implies proportionality.
\end{theorem}

\begin{proof}
Consider a matching instance and let $\x$ be an envy-free lottery over matchings. Abbreviate the rank requirement for agent $i\in N$ for proportionality by $p_i=\left\lceil n(1-h_i)\right\rceil$ and assume that $\x$ is not proportional. Then, there exists an agent $i^*\in N$ such that
    \begin{align*}
        \rank_{i^*}(\rep(x_{i^*},\succ_{i^*},h_{i^*})) &\geq p_{i^*}+1.
    \end{align*}
    Since $\x$ is envy-free, we have
    \begin{align}\label{eq:ef}
        \rank_{i^*}(\rep(x_i,\succ_{i^*},h_{i^*})) &\geq \rank_{i^*}(\rep(x_{i^*},\succ_{i^*},h_{i^*})) \geq p_{i^*}+1,
    \end{align}
    for every agent $i\in N$. Denote by $g^*$ the item ranked $p_{i^*}$-th by agent ${i^*}$. By the definition of $\rep(x_i,\succ_{i^*},h_{i^*})$ and inequality (\ref{eq:ef}), we have
    \begin{align*}
        \sum_{g\in M:g^*\succ_{i^*} g}{x_{ig}} &> h_{i^*}
    \end{align*}
    for every agent $i\in N$. Using this inequality, we have
    \begin{align}\label{eq:double-sum}
        \sum_{g\in M:g^*\succ_{i^*} g}{\sum_{i\in N}{x_{ig}}} & =\sum_{i\in N}{\sum_{g\in M:g^*\succ_{i^*} g}{x_{ig}}} > n\cdot h_{i^*}.
    \end{align}
    Now, notice that there are 
    \begin{align*}
        n-p_{i^*} =n-\lceil n(1-h_{i^*})\rceil \leq n\cdot h_{i^*}
    \end{align*}
    items ranked below the item $g^*$ by agent $i^*$ and, hence, at most $n\cdot h_{i^*}$ terms in the outer sum of the LHS of Equation~(\ref{eq:double-sum}). Thus, there exists one such term corresponding to an item $g'\in M$ such that $\sum_{i\in N}{x_{ig'}}>1$, violating the validity of lottery $\x$.
\end{proof}

Unfortunately, envy-freeness and efficiency may not be possible simultaneously, as the next simple example illustrates.
\begin{example}
Consider a one-sided matching instance with two agents $1$ and $2$ with identical preference $a\succ b$ for two items $a$ and $b$ and identical quantile parameter $h<1/2$. First, observe that the same item cannot become the representative of both agents. If item $a$ is the representative for both agents, this means that $x_{1a}\geq 1-h>1/2$ and $x_{2a}\geq 1-h>1/2$, violating the matching constraint $x_{1a}+x_{2a}\leq 1$. The lottery that makes item $b$ representative for both agents (e.g., the uniform lottery) is dominated by the lotteries that make item $a$ representative for agent $1$ and item $b$ representative for agent $2$ (e.g., $x_{1a}=x_{2b}=1$ and $x_{1b}=x_{2a}=0$) and vice-versa. These two classes of efficient lotteries are clearly not envy-free. \qed
\end{example}

Therefore, the following question arises as a very natural one.
\begin{question}
    What is the complexity of deciding whether an efficient and envy-free lottery exists for a given one-sided matching instance with quantile agent utilities?
\end{question}

Nevertheless, we show that a modification of the SDP mechanism, which inherits its efficiency, strategyproofness, and proportionality, returns envy-free lotteries provided that the agents have {\em different} quantile parameters.

\paragraph{Serial dictatorship for envy-freeness (SDEF)} First, execute the SDP mechanism to compute $\LP(\profile,\rr)$. Then, further constrain the linear program $\LP(\profile,\rr)$ by considering the agents in reverse order (i.e., in non-decreasing order in terms of their quantile parameters). When considering agent $i$, add additional constraints to $\LP(\profile,\rr)$ which maximize the probability that agent $i$ gets her most preferred item, subject to this, maximize the probability that the agent gets her second most preferred item, so so on, so that $\LP(\profile,\rr)$ remains feasible. After all agents have been considered, mechanism SDEF returns an arbitrary lottery over matchings that satisfies $\LP(\profile; \rr)$.

The proof of the next statement appears in Appendix~\ref{app:sec:omitted-proofs}.
\begin{restatable}{theorem}{thmonesidesdef}\label{thm:SDEF}
    Mechanism SDEF is efficient, strategyproof, proportional, and envy-free
    if $h_i \neq h_j$ for all $i \neq j \in N$. 
\end{restatable}

We conclude this section by emphasizing the necessity of the reverse phase in the definition of mechanism SDEF. Without it, the lottery returned by mechanism SDP may not be envy-free as the next example illustrates.
\begin{example}
Consider the one-sided matching instance with two agents having the same preference $a \succ b$ and quantile parameters $h_1  = 0.3$ and $h_2 = 0.2$. By executing the SDP mechanism, we get $a$ as the representative of agent $1$, and $b$ as the representative of agent $2$. The lottery $\x'$ with $x'_{1a} = 1 = x'_{2b}$ would be a valid solution returned by mechanism SDP. However, agent $2$ would envy agent $1$, since she has her more preferred item $a$ as representative under lottery $x'_1$. The lottery $\x$ returned by mechanism SDEF has instead $x_{1a} = 0.7$ and $x_{1b} = 0.3$, which guarantees envy-freeness. Agent $2$ has item $b$ as representative and would still do so under lottery $x_1$.
\end{example}

\section{Conclusion}\label{sec:conclusion} 
We have initiated a novel direction in randomized social choice by proposing the new concept of quantile agent utility for randomized outcomes. We discussed its advantages compared to other well-studied utility models and presented a series of results that demonstrate the power of the new definition in various social choice settings, including voting and one-sided matchings, as well as its implications for concepts such as efficiency, strategyproofness, and fairness. We have identified several open problems, suggesting interesting directions for future work. 

One direction we have not considered here is related to a stronger notion of strategyproofness. Imagine the scenario in which the quantile parameters are not known a priori (as we assume in the current work) but instead are reported by the agents together with their preferences to the mechanism. Are there mechanisms (i.e., voting rules or matching mechanisms) that are robust to manipulation by agents who can misreport both their preference and quantile parameter? We can show that mechanism SD is strategyproof under this stronger notion. However, the question of whether it is compatible with efficiency and fairness deserves further investigation.

\bibliographystyle{ACM-Reference-Form}
\bibliography{arxivBib}


\begin{thebibliography}{29}


\ifx \showCODEN    \undefined \def \showCODEN     #1{\unskip}     \fi
\ifx \showDOI      \undefined \def \showDOI       #1{#1}\fi
\ifx \showISBNx    \undefined \def \showISBNx     #1{\unskip}     \fi
\ifx \showISBNxiii \undefined \def \showISBNxiii  #1{\unskip}     \fi
\ifx \showISSN     \undefined \def \showISSN      #1{\unskip}     \fi
\ifx \showLCCN     \undefined \def \showLCCN      #1{\unskip}     \fi
\ifx \shownote     \undefined \def \shownote      #1{#1}          \fi
\ifx \showarticletitle \undefined \def \showarticletitle #1{#1}   \fi
\ifx \showURL      \undefined \def \showURL       {\relax}        \fi
\providecommand\bibfield[2]{#2}
\providecommand\bibinfo[2]{#2}
\providecommand\natexlab[1]{#1}
\providecommand\showeprint[2][]{arXiv:#2}

\bibitem[Abdulkadiro{\u{g}}lu and S{\"o}nmez(2003)]%
        {abdulkadirouglu2003ordinal}
\bibfield{author}{\bibinfo{person}{Atila Abdulkadiro{\u{g}}lu} {and}
  \bibinfo{person}{Tayfun S{\"o}nmez}.} \bibinfo{year}{2003}\natexlab{}.
\newblock \showarticletitle{Ordinal efficiency and dominated sets of
  assignments}.
\newblock \bibinfo{journal}{\emph{Journal of Economic Theory}}
  \bibinfo{volume}{112}, \bibinfo{number}{1} (\bibinfo{year}{2003}),
  \bibinfo{pages}{157--172}.
\newblock


\bibitem[Abdulkadiro\u{g}lu and S\"{o}nmez(1998)]%
        {AS98}
\bibfield{author}{\bibinfo{person}{Atila Abdulkadiro\u{g}lu} {and}
  \bibinfo{person}{Tayfun S\"{o}nmez}.} \bibinfo{year}{1998}\natexlab{}.
\newblock \showarticletitle{Random Serial Dictatorship and the Core from Random
  Endowments in House Allocation Problems}.
\newblock \bibinfo{journal}{\emph{Econometrica}} \bibinfo{volume}{66},
  \bibinfo{number}{3} (\bibinfo{year}{1998}), \bibinfo{pages}{689--701}.
\newblock


\bibitem[Arrow(1951)]%
        {arrow1951}
\bibfield{author}{\bibinfo{person}{Kenneth~J. Arrow}.}
  \bibinfo{year}{1951}\natexlab{}.
\newblock \bibinfo{booktitle}{\emph{Social Choice and Individual Values}
  (\bibinfo{edition}{2nd edition, 1963} ed.)}.
\newblock \bibinfo{publisher}{John Wiley \& Sons}.
\newblock


\bibitem[Aziz(2019)]%
        {aziz2019probabilistic}
\bibfield{author}{\bibinfo{person}{Haris Aziz}.}
  \bibinfo{year}{2019}\natexlab{}.
\newblock \showarticletitle{A probabilistic approach to voting, allocation,
  matching, and coalition formation}.
\newblock In \bibinfo{booktitle}{\emph{The Future of Economic Design}},
  \bibfield{editor}{\bibinfo{person}{Jean-François Laslier},
  \bibinfo{person}{Hervé Moulin}, \bibinfo{person}{M.~Remzi Sanver}, {and}
  \bibinfo{person}{William~S. Zwicker}} (Eds.). \bibinfo{publisher}{Springer},
  \bibinfo{pages}{45--50}.
\newblock


\bibitem[Aziz et~al\mbox{.}(2019)]%
        {ABL+19}
\bibfield{author}{\bibinfo{person}{Haris Aziz}, \bibinfo{person}{Péter Biró},
  \bibinfo{person}{Jérôme Lang}, \bibinfo{person}{Julien Lesca}, {and}
  \bibinfo{person}{Jérôme Monnot}.} \bibinfo{year}{2019}\natexlab{}.
\newblock \showarticletitle{Efficient reallocation under additive and
  responsive preferences}.
\newblock \bibinfo{journal}{\emph{Theoretical Computer Science}}
  \bibinfo{volume}{790} (\bibinfo{year}{2019}), \bibinfo{pages}{1--15}.
\newblock


\bibitem[Aziz et~al\mbox{.}(2018)]%
        {a18}
\bibfield{author}{\bibinfo{person}{Haris Aziz}, \bibinfo{person}{Florian
  Brandl}, \bibinfo{person}{Felix Brandt}, {and} \bibinfo{person}{Markus
  Brill}.} \bibinfo{year}{2018}\natexlab{}.
\newblock \showarticletitle{On the tradeoff between efficiency and
  strategyproofness}.
\newblock \bibinfo{journal}{\emph{Games and Economic Behavior}}
  \bibinfo{volume}{110} (\bibinfo{year}{2018}), \bibinfo{pages}{1--18}.
\newblock


\bibitem[Bogomolnaia and Moulin(2001)]%
        {bogomolnaia2001}
\bibfield{author}{\bibinfo{person}{Anna Bogomolnaia} {and}
  \bibinfo{person}{Herv{\'e} Moulin}.} \bibinfo{year}{2001}\natexlab{}.
\newblock \showarticletitle{A New Solution to the Random Assignment Problem}.
\newblock \bibinfo{journal}{\emph{Journal of Economic Theory}}
  \bibinfo{volume}{100}, \bibinfo{number}{2} (\bibinfo{year}{2001}),
  \bibinfo{pages}{295--328}.
\newblock


\bibitem[Brams and Taylor(1996)]%
        {BT96}
\bibfield{author}{\bibinfo{person}{Steven~J. Brams} {and}
  \bibinfo{person}{Alan~D. Taylor}.} \bibinfo{year}{1996}\natexlab{}.
\newblock \bibinfo{booktitle}{\emph{Fair Division: From Cake-Cutting to Dispute
  Resolution}}.
\newblock \bibinfo{publisher}{Cambridge University Press}.
\newblock


\bibitem[Brandt(2017)]%
        {B17}
\bibfield{author}{\bibinfo{person}{Felix Brandt}.}
  \bibinfo{year}{2017}\natexlab{}.
\newblock \showarticletitle{Rolling the dice: Recent results in probabilistic
  social choice}.
\newblock In \bibinfo{booktitle}{\emph{Trends in Computational Social Choice}},
  \bibfield{editor}{\bibinfo{person}{Ulle Endriss}} (Ed.).
  \bibinfo{publisher}{AI Access}, \bibinfo{pages}{3--26}.
\newblock


\bibitem[Caragiannis et~al\mbox{.}(2021)]%
        {caragiannis2021interim}
\bibfield{author}{\bibinfo{person}{Ioannis Caragiannis},
  \bibinfo{person}{Panagiotis Kanellopoulos}, {and} \bibinfo{person}{Maria
  Kyropoulou}.} \bibinfo{year}{2021}\natexlab{}.
\newblock \showarticletitle{On interim envy-free allocation lotteries}. In
  \bibinfo{booktitle}{\emph{Proceedings of the 22nd ACM Conference on Economics
  and Computation (EC)}}. \bibinfo{pages}{264--284}.
\newblock


\bibitem[Emanuel et~al\mbox{.}(2020)]%
        {NEJM20}
\bibfield{author}{\bibinfo{person}{Ezekiel~J. Emanuel}, \bibinfo{person}{Govind
  Persad}, \bibinfo{person}{Ross Upshur}, \bibinfo{person}{Beatriz Thome},
  \bibinfo{person}{Michael Parker}, \bibinfo{person}{Aaron Glickman},
  \bibinfo{person}{Cathy Zhang}, \bibinfo{person}{Connor Boyle},
  \bibinfo{person}{Maxwell Smith}, {and} \bibinfo{person}{James~P. Phillips}.}
  \bibinfo{year}{2020}\natexlab{}.
\newblock \showarticletitle{Fair Allocation of Scarce Medical Resources in the
  Time of Covid-19}.
\newblock \bibinfo{journal}{\emph{New England Journal of Medicine}}
  \bibinfo{volume}{382}, \bibinfo{number}{21} (\bibinfo{year}{2020}),
  \bibinfo{pages}{2049--2055}.
\newblock


\bibitem[Fishburn(1972)]%
        {fishburn1972lotteries}
\bibfield{author}{\bibinfo{person}{Peter~C. Fishburn}.}
  \bibinfo{year}{1972}\natexlab{}.
\newblock \showarticletitle{Lotteries and social choices}.
\newblock \bibinfo{journal}{\emph{Journal of Economic Theory}}
  \bibinfo{volume}{5}, \bibinfo{number}{2} (\bibinfo{year}{1972}),
  \bibinfo{pages}{189--207}.
\newblock


\bibitem[Flanigan et~al\mbox{.}(2021)]%
        {Flanigan2021FairAssemblies}
\bibfield{author}{\bibinfo{person}{Bailey Flanigan}, \bibinfo{person}{Paul
  G{\"o}lz1}, \bibinfo{person}{Anupam Gupta}, \bibinfo{person}{Brett Hennig},
  {and} \bibinfo{person}{Ariel~D. Procaccia}.} \bibinfo{year}{2021}\natexlab{}.
\newblock \showarticletitle{Fair algorithms for selecting citizens’
  assemblies}.
\newblock \bibinfo{journal}{\emph{Nature}} \bibinfo{volume}{596},
  \bibinfo{number}{7873} (\bibinfo{year}{2021}), \bibinfo{pages}{548--552}.
\newblock
\urldef\tempurl%
\url{https://doi.org/10.1038/s41586-021-03788-6}
\showDOI{\tempurl}


\bibitem[Foley(1966)]%
        {foley1966resource}
\bibfield{author}{\bibinfo{person}{Duncan~Karl Foley}.}
  \bibinfo{year}{1966}\natexlab{}.
\newblock \bibinfo{booktitle}{\emph{Resource allocation and the public
  sector}}.
\newblock \bibinfo{publisher}{Yale University}.
\newblock


\bibitem[Gibbard(1973)]%
        {Gib73}
\bibfield{author}{\bibinfo{person}{Allan Gibbard}.}
  \bibinfo{year}{1973}\natexlab{}.
\newblock \showarticletitle{Manipulation of Voting Schemes: A General Result}.
\newblock \bibinfo{journal}{\emph{Econometrica}} \bibinfo{volume}{41},
  \bibinfo{number}{4} (\bibinfo{year}{1973}), \bibinfo{pages}{587--602}.
\newblock


\bibitem[Gibbard(1977)]%
        {G77chance}
\bibfield{author}{\bibinfo{person}{Allan Gibbard}.}
  \bibinfo{year}{1977}\natexlab{}.
\newblock \showarticletitle{Manipulation of Schemes that Mix Voting with
  Chance}.
\newblock \bibinfo{journal}{\emph{Econometrica}} \bibinfo{volume}{45},
  \bibinfo{number}{3} (\bibinfo{year}{1977}), \bibinfo{pages}{665--681}.
\newblock


\bibitem[Hylland and Zeckhauser(1979)]%
        {hylland1979efficient}
\bibfield{author}{\bibinfo{person}{Aanund Hylland} {and}
  \bibinfo{person}{Richard Zeckhauser}.} \bibinfo{year}{1979}\natexlab{}.
\newblock \showarticletitle{The efficient allocation of individuals to
  positions}.
\newblock \bibinfo{journal}{\emph{Journal of Political Economy}}
  \bibinfo{volume}{87}, \bibinfo{number}{2} (\bibinfo{year}{1979}),
  \bibinfo{pages}{293--314}.
\newblock


\bibitem[Intriligator(1973)]%
        {intriligator1973probabilistic}
\bibfield{author}{\bibinfo{person}{Michael~D. Intriligator}.}
  \bibinfo{year}{1973}\natexlab{}.
\newblock \showarticletitle{A probabilistic model of social choice}.
\newblock \bibinfo{journal}{\emph{The Review of Economic Studies}}
  \bibinfo{volume}{40}, \bibinfo{number}{4} (\bibinfo{year}{1973}),
  \bibinfo{pages}{553--560}.
\newblock


\bibitem[Karp(1972)]%
        {Karp72a}
\bibfield{author}{\bibinfo{person}{Richard~M. Karp}.}
  \bibinfo{year}{1972}\natexlab{}.
\newblock \showarticletitle{Reducibility among Combinatorial Problems}.
\newblock In \bibinfo{booktitle}{\emph{Complexity of Computer Computations}},
  \bibfield{editor}{\bibinfo{person}{Raymond~E. Miller} {and}
  \bibinfo{person}{James~W. Thatcher}} (Eds.). \bibinfo{publisher}{Plenum
  Press}, \bibinfo{pages}{85--103}.
\newblock


\bibitem[Manski(1988)]%
        {manski1988ordinal}
\bibfield{author}{\bibinfo{person}{Charles~F Manski}.}
  \bibinfo{year}{1988}\natexlab{}.
\newblock \showarticletitle{Ordinal utility models of decision making under
  uncertainty}.
\newblock \bibinfo{journal}{\emph{Theory and Decision}} \bibinfo{volume}{25},
  \bibinfo{number}{1} (\bibinfo{year}{1988}), \bibinfo{pages}{79--104}.
\newblock


\bibitem[Procaccia(2010)]%
        {procaccia2010}
\bibfield{author}{\bibinfo{person}{Ariel~D. Procaccia}.}
  \bibinfo{year}{2010}\natexlab{}.
\newblock \showarticletitle{Can Approximation Circumvent
  {G}ibbard-{S}atterthwaite?}. In \bibinfo{booktitle}{\emph{Proceedings of the
  26th Conference on Uncertainty in Artificial Intelligence (UAI)}}.
  \bibinfo{pages}{597--606}.
\newblock


\bibitem[Rostek(2010)]%
        {rostek2010quantile}
\bibfield{author}{\bibinfo{person}{Marzena Rostek}.}
  \bibinfo{year}{2010}\natexlab{}.
\newblock \showarticletitle{Quantile maximization in decision theory}.
\newblock \bibinfo{journal}{\emph{The Review of Economic Studies}}
  \bibinfo{volume}{77}, \bibinfo{number}{1} (\bibinfo{year}{2010}),
  \bibinfo{pages}{339--371}.
\newblock


\bibitem[Satterthwaite(1975)]%
        {Sat75}
\bibfield{author}{\bibinfo{person}{Mark~Allen Satterthwaite}.}
  \bibinfo{year}{1975}\natexlab{}.
\newblock \showarticletitle{Strategy-proofness and {A}rrow's conditions:
  Existence and correspondence theorems for voting procedures and social
  welfare functions}.
\newblock \bibinfo{journal}{\emph{Journal of Economic Theory}}
  \bibinfo{volume}{10}, \bibinfo{number}{2} (\bibinfo{year}{1975}),
  \bibinfo{pages}{187--217}.
\newblock


\bibitem[Sen(1970)]%
        {Sen1970CCSW}
\bibfield{author}{\bibinfo{person}{Amartya~K. Sen}.}
  \bibinfo{year}{1970}\natexlab{}.
\newblock \bibinfo{booktitle}{\emph{Collective Choice and Social Welfare}}.
\newblock \bibinfo{publisher}{Holden-Day}, \bibinfo{address}{San Francisco}.
\newblock


\bibitem[Shapley and Scarf(1974)]%
        {SS74}
\bibfield{author}{\bibinfo{person}{Lloyd Shapley} {and}
  \bibinfo{person}{Herbert Scarf}.} \bibinfo{year}{1974}\natexlab{}.
\newblock \showarticletitle{On cores and indivisibility}.
\newblock \bibinfo{journal}{\emph{Journal of Mathematical Economics}}
  \bibinfo{volume}{1}, \bibinfo{number}{1} (\bibinfo{year}{1974}),
  \bibinfo{pages}{23--37}.
\newblock


\bibitem[Steinhaus(1948)]%
        {steinhaus1948problem}
\bibfield{author}{\bibinfo{person}{Hugo Steinhaus}.}
  \bibinfo{year}{1948}\natexlab{}.
\newblock \showarticletitle{The problem of fair division}.
\newblock \bibinfo{journal}{\emph{Econometrica}}  \bibinfo{volume}{16}
  (\bibinfo{year}{1948}), \bibinfo{pages}{101--104}.
\newblock


\bibitem[Varian(1974)]%
        {varian1973equity}
\bibfield{author}{\bibinfo{person}{Hal~R. Varian}.}
  \bibinfo{year}{1974}\natexlab{}.
\newblock \showarticletitle{Equity, envy, and efficiency}.
\newblock \bibinfo{journal}{\emph{Journal of Economic Theory}}
  \bibinfo{volume}{9}, \bibinfo{number}{1} (\bibinfo{year}{1974}),
  \bibinfo{pages}{63--91}.
\newblock


\bibitem[White et~al\mbox{.}(2022)]%
        {White22}
\bibfield{author}{\bibinfo{person}{Douglas~B. White}, \bibinfo{person}{Erin~K.
  McCreary}, \bibinfo{person}{Chung-Chou~H. Chang}, \bibinfo{person}{Mark
  Schmidhofer}, \bibinfo{person}{J.~Ryan Bariola}, \bibinfo{person}{Naudia~N.
  Jonassaint}, \bibinfo{person}{Govind Persad}, \bibinfo{person}{Robert~D.
  Truog}, \bibinfo{person}{Parag Pathak}, \bibinfo{person}{Tayfun Sonmez},
  {and} \bibinfo{person}{M.~Utku Unver}.} \bibinfo{year}{2022}\natexlab{}.
\newblock \showarticletitle{A Multicenter Weighted Lottery to Equitably
  Allocate Scarce COVID-19 Therapeutics}.
\newblock \bibinfo{journal}{\emph{American Journal of Respiratory and Critical
  Care Medicine}} \bibinfo{volume}{206}, \bibinfo{number}{4}
  (\bibinfo{year}{2022}), \bibinfo{pages}{503--506}.
\newblock


\bibitem[Zeckhauser(1969)]%
        {zeckhauser1969majority}
\bibfield{author}{\bibinfo{person}{Richard Zeckhauser}.}
  \bibinfo{year}{1969}\natexlab{}.
\newblock \showarticletitle{Majority rule with lotteries on alternatives}.
\newblock \bibinfo{journal}{\emph{The Quarterly Journal of Economics}}
  \bibinfo{volume}{83}, \bibinfo{number}{4} (\bibinfo{year}{1969}),
  \bibinfo{pages}{696--703}.
\newblock


\end{thebibliography}

\appendix

\section{Omitted proofs}\label{app:sec:omitted-proofs}
\thmonesidenp*
\begin{proof}
In the following, we give a reduction from vertex cover, which is one of Karp's 21 complete problems \cite{Karp72a}. 
Given a vertex cover instance $G = (V,E,k)$,
we construct the following one-sided matching instance.
Let $N = A_V \cup A_E$ be the set of agents with $A_V = \{a_v \mid v \in V\}$ and $A_E = \{a_e \mid e\in E\}$ and let $M = V \cup E$ and  $\lvert M \rvert = m$. 
Furthermore, we set $r,h$ as follows:
    \begin{itemize}
        \item For every $v \in V$, set for agent $a_v$ $r_{a_v} = 1$ and $h_{a_v}= \frac{1}{3m}$. Furthermore, let her top choice be $v$ (so the only way to satisfy $a_v$ is to put at least $1- \frac{1}{3m}$ probability on $v$).
        \item For each edge $e =\{u,v\} \in E$, we set $r_{a_e} = 2$  and $h_{a_e} = 1-\frac{1}{m}$, and  the top two choices of $a_e$ are $u,v$ (so, to satisfy $a_e$, by the pigeonhole principle, at least $\frac{1}{2m}$ probability has to be either on $u$ or on $v$). 
    \end{itemize}
    We claim that there exists a vertex cover of size $k$ if and only if there is a solution in which $\lvert V \rvert +\lvert E \rvert-k$ agents are satisfied.

    So, let $C$ be a vertex cover of size $k$. Then, we can define the following partial probability distribution $x$. If $u \in C$, then for all $e$ such that $u \in e$, we set $x_{a_eu} = \frac{1}{m}$. Furthermore, for each $v \not \in C$, we set $x_{a_vv} = 1$. Trivially, every agent $a_v$ with $v \not \in C$ is satisfied. Furthermore, since $C$ is a vertex cover for every agent $a_e$ with $e = \{u,v\}$ it holds that $x_{a_eu} + x_{a_ev} \geq \frac{1}{m}$. Thus, every agent $a_e$ is satisfied. In total this means that at least $\lvert V\rvert + \lvert E \rvert - k$ agents are satisfied.

    Let $x$ be a probability distribution such that a set $S$ of $\lvert V\rvert + \lvert E \rvert - k$ agents are satisfied. We claim that we can construct a vertex cover $C$ of size at most $k$ as follows:
    $C = V' \cup  \bigcup_{a_{\{u,v\}} \in E'} \{f(\{u,v\})\}$ where $V' = \{v \mid x_{a_v,v} < 1- \frac{1}{3m}\}$, $E' =\{a_e \mid e \in E \} \cap (N \setminus S)$, and $f: E \rightarrow V$ is any function that maps every edge to one of its two incident vertices.
    Observe that every agent $a_v$ with $v \in V'$ and every agent in $E'$ is unsatisfied. Thus, $\lvert C \rvert \leq k$. It remains to show that $C$ forms a vertex cover. Let $e = \{u,v\}$ be any edge in $E$. If $u$ or $v$ in $V'$, then $e$ is covered. So, assume that $e \cap V' = \emptyset$.
    Thus, it holds that $x_{a_eu} < \frac{1}{3m}$ and $x_{a_ev} < \frac{1}{3m}$ and therefore $x_{a_eu} + x_{a_ev} < \frac{1}{m}$. This implies that $a_e \not\in S$ and, therefore, $a_e \in E'$. By definition, we get that $f(e)$ is in $C$ and thus $C \cap e \not = \emptyset$. 
\end{proof}

\thmonesidesdef*
\begin{proof}
We use the terms Phase 0, Phase 1, and Phase 2 for the application of mechanism SDP, the tightening of the linear program by considering the agents in reverse order, and the selection of the final lottery, respectively. We divide the proof into three parts. We first show that the probability $\x$ distributed in Phase $1$ is sufficient to guarantee each agent their representative (Claim 1).
Secondly, we prove that if an agent $i$ gets representative $c_i$, then for all $c \succ_i c_i$, it holds that 
$\sum_{j \in N} x_{jc} = 1$ after Phase 1 (Claim 2). Finally, we show that no agent envies another agent. 

First we prove Claim 1.
Consider agent $i$'s turn in Phase $1$. We distribute as much probability as possible on her top choice, then on her second choice, and so on ,while upholding the ranks determined by Phase 0. We claim that for no agent $i$ we distribute any positive probability on any item $c$ such that $c_i \succ c$. Observe that, for agent $i$, we have rank requirement $\sum_{c \succsim c_i} x_{ic} \geq 1-h_i$. By definition, there exists a distribution that extends the already assigned probabilities to a lottery while satisfying the rank requirement for every agent and therefore also for agent $i$. 
When agent $i$ is considered in Phase 1, we put as much probability as possible on her top choice 
without violating the rank requirements. So, after distributing that probability, there is still an extension of the currently distributed probability that upholds the rank requirement. Thus, we do not distribute any probability on 
any item less preferred than $c_i$, thus proving Claim 1. 

Next, we prove Claim 2. Assume for the sake of contradiction that the claim does not hold. Then, there exists an agent $i$ such that $\sum_j x_{jc} < 1$ after Phase 1 for some item $c \succ_i c_i$. Furthermore, let $c$ to be the most preferred item of agent $i$ with $\sum_j x_{jc} < 1$ and let $\epsilon > 0$ be any rational number such that $\sum_j x_{jc} + \epsilon < 1$ as well as $\epsilon < x_{ic_i}$.
We can now find another distribution $x'$ in which $x'_{ic} = x_{ic} + \epsilon$ and $x'_{ic_i} = x_{ic_i} - \epsilon$. This distribution will give each agent the same ranks as $x$ (we only changed the assignment of $i$) but puts more probability on more preferred alternatives of agent $i$ which contradicts the fact that in Phase $1$ the mechanism puts as much probability as possible on the best choices of agent $i$.

We can now prove that the mechanism is envy-free.
Let $i$ and $j$ be any agents with $h_i > h_j$. We will first prove that $i$ does not envy $j$. Let $c_i = \rep(x_i,\succ_i,h_i)$ and $c_j = \rep(x_j,\succ_i,h_i)$. Since $h_i > h_j$, it holds that agent $i$ is considered first in the mechanism and receives the best possible representative without making any previously considered agent worse off. Since $j$
is considered later, agent $i$ cannot envy her. This follows since if agent $i$ envies her, consider the following solution $x'$ for the linear program when considering agent $i$: $x'_k = x_k$ for all $k \neq \{i,j\}$, $x_i' = x_j$ and $x_j' = x_i$. Then $x'$ is a solution that gives every agent except $i$ and $j$ the same but improves the rank of agent $i$. We have reached a contradiction.

It remains to show that agent $j$ does not envy agent $i$. 
Let $c_i = \rep(x_i,\succ_i,h_j)$ and $c_j = \rep(x_j,\succ_i,h_i)$.
Since, in Phase 1, a probability of at most $1-h_i < 1-h_j$ was distributed, it holds that a probability of $h_i- h_j$  was distributed in Phase 2 on items which agent $j$ weakly prefers to $c_i$. By Claim 2, this holds for all $c \succ_j c_j$ to which all probability is distributed in Phase 1. Therefore, $c_j \succsim_j c_i$ proving that $j$ has no envy towards $i$. 
\end{proof}

\end{document}